\documentclass[sigconf]{acmart}

\renewcommand\footnotetextcopyrightpermission[1]{} 
\AtBeginDocument{%
}

\setcopyright{acmlicensed}
\copyrightyear{2025}
\acmYear{2025}
\acmDOI{XXXXXXX.XXXXXXX}
\acmISBN{978-1-4503-XXXX-X/2018/06}

\usepackage{tikz}
\usepackage{amsmath}
\usepackage{pifont}
\usepackage{makecell}
\usepackage[most]{tcolorbox}
\usepackage[normalem]{ulem}
\usepackage{shadowtext}
\usepackage{censor}
\usepackage{caption}
\usepackage{xspace}
\usepackage{url}            
\usepackage{booktabs}       
\usepackage{nicefrac}       
\usepackage{microtype}      
\usepackage{setspace}
\usepackage[boxed, ruled, vlined, linesnumbered]{algorithm2e}
\usepackage{algorithmic}
\usepackage{wrapfig}
\usepackage{graphicx}
\usepackage{multirow}
\usepackage{colortbl}
\usepackage{soul}
\usepackage[colorinlistoftodos]{todonotes}
\usepackage{multicol}
\usepackage{xcolor}  
\usepackage{pdfpages}
\usepackage{marvosym}
\usepackage{hyperref}       
\usepackage{cleveref}

\newenvironment{changemargin}[2]{\begin{list}{}{
	\setlength{\topsep}{0pt}\setlength{\leftmargin}{11pt}
	\setlength{\rightmargin}{0pt}
	\setlength{\listparindent}{\parindent}
	\setlength{\itemindent}{\parindent}
	\setlength{\parsep}{0pt plus 1pt}
	\addtolength{\leftmargin}{#1}\addtolength{\rightmargin}{#2}
	}\item}
	{\end{list}}
    
\newenvironment{mitemize}{
	\begin{changemargin}{-10pt}{-0cm}
	\vspace{-10pt}
	\hspace{-8pt}
	\begin{itemize}
	\setlength{\itemsep}{3pt}}
	{\end{itemize}
	\vspace{2pt}
	\end{changemargin}}

\newcommand{\system}{\textsc{CyLens}\xspace}
\newcommand{\systems}{\textsc{{\footnotesize CyLens}}\xspace}
\newcommand{\sevenllm}{\textsc{{\footnotesize SEvenLLM}}\xspace}
\newcommand{\lily}{\textsc{{\footnotesize Lily-Cyber}}\xspace}
\newcommand{\gpt}{\textsc{{\footnotesize ChatGPT-4o}}\xspace}
\newcommand{\gpto}{\textsc{{\footnotesize ChatGPT-o1}}\xspace}
\newcommand{\gemini}{\textsc{{\footnotesize Gemini-Pro}}\xspace}
\newcommand{\llama}{\textsc{{\footnotesize Llama-405B}}\xspace}
\newcommand{\deepseek}{\textsc{{\footnotesize DeepSeek-R1}}\xspace}
\newcommand{\claude}{\textsc{{\footnotesize Claude-Opus}}\xspace}

\newcommand{\ncve}{271,570\xspace}

\definecolor{darkgreen}{rgb}{0, 0.5, 0}  

\newcommand{\cmark}{\textcolor{darkgreen}{\ding{51}}} 

\definecolor{cellcolor}{RGB}{0,0,0}
\definecolor{antiquewhite}{rgb}{0.98, 0.92, 0.84}
\definecolor{anti-flashwhite}{rgb}{0.95, 0.95, 0.96}
\definecolor{aliceblue}{rgb}{0.94, 0.97, 1.0}
\definecolor{almond}{rgb}{0.94, 0.87, 0.8}
\definecolor{cosmiclatte}{rgb}{1.0, 0.97, 0.91}
\definecolor{darkbyzantium}{rgb}{0.36, 0.22, 0.33}
\definecolor{darkseagreen}{rgb}{0.56, 0.74, 0.56}
\definecolor{darkspringgreen}{rgb}{0.09, 0.45, 0.27}
\definecolor{asparagus}{rgb}{0.53, 0.66, 0.42}
\definecolor{antiquefuchsia}{rgb}{0.57, 0.36, 0.51}
\definecolor{ao(english)}{rgb}{0.0, 0.5, 0.0}
\definecolor{deepcerise}{rgb}{0.85, 0.2, 0.53}
\definecolor{denim}{rgb}{0.08, 0.38, 0.74}
\definecolor{crimson}{rgb}{0.86, 0.08, 0.24}
\definecolor{buff}{rgb}{0.94, 0.86, 0.51}
\definecolor{amber(sae/ece)}{rgb}{1.0, 0.49, 0.0}
\definecolor{airforceblue}{rgb}{0.36, 0.54, 0.66}
\definecolor{amethyst}{rgb}{0.6, 0.4, 0.8}
\definecolor{azure(colorwheel)}{rgb}{0.0, 0.5, 1.0}
\definecolor{azure(web)(azuremist)}{rgb}{0.94, 1.0, 1.0}
\definecolor{beige}{rgb}{0.96, 0.96, 0.86}
\definecolor{cornsilk}{rgb}{1.0, 0.97, 0.86}
\definecolor{cosmiclatte}{rgb}{1.0, 0.97, 0.91}

\newcommand{\hlcell}{\cellcolor{denim!15}}
\newcommand{\boxcolor}{cosmiclatte!10}

\newtheorem{example}{Example}
\newtheorem{case}{Case Study}
\newtheorem{prompt}{Prompt}

\begin{document}

\title{\system: Towards Reinventing Cyber Threat Intelligence in the Paradigm of Agentic Large Language Models}



\author{Xiaoqun Liu}
\affiliation{
\institution{Stony Brook University}
\country{Stony Brook, NY, USA}
}
\email{xiaoqun.will@gmail.com}

\author{Jiacheng Liang\textsuperscript{*}}
\thanks{\textsuperscript{*} Xiaoqun Liu and Jiacheng Liang contributed equally to this work.}
\affiliation{%
\institution{Stony Brook University}
\country{Stony Brook, NY, USA}
}
\email{ljcpro@outlook.com}

\author{Qiben Yan}
\affiliation{%
\institution{Michigan State University}
\country{East Lansing, MI, USA}
}
\email{qyan@msu.edu}

\author{Jiyong Jang}
\affiliation{%
\institution{IBM Research}
\country{Yorktown Heights, NY, USA}
}
\email{jjang@us.ibm.com}

\author{Sicheng Mao}
\affiliation{%
\institution{Google}
\country{New York City, NY, USA}
}
\email{sichengm7@gmail.com}

\author{Muchao Ye}
\affiliation{%
\institution{The University of Iowa}
\country{Iowa City, IA, USA}
}
\email{muchao-ye@uiowa.edu}

\author{Jinyuan Jia}
\affiliation{%
\institution{Pennsylvania State University}
\country{State College, PA, USA}
}
\email{jinyuan@psu.edu}

\author{Zhaohan Xi}
\affiliation{%
\institution{Binghamton University}
\country{Vestal, NY, USA}
}
\email{zxi1@binghamton.edu}

\begin{abstract}

The exponential growth of cyber threat knowledge, exemplified by the expansion of databases such as MITRE-CVE and NVD, poses significant challenges for cyber threat analysis. Security professionals are increasingly burdened by the sheer volume and complexity of information, creating an urgent need for effective tools to synthesize massive knowledge and counter evolving threats proactively. However, conventional threat intelligence often struggle to scale with the rapidly expanding threat landscape and lack the adaptability required to support diverse threat intelligence tasks.

In this work, we introduce \system, a cyber threat intelligence system powered by large language models (LLMs). \system is designed to assist security professionals throughout the entire threat analysis lifecycle, supporting threat attribution, contextualization, correlation, prioritization, and remediation. To ensure domain expertise, \system integrates knowledge from \ncve threat reports into its model parameters and incorporates six specialized NLP modules to enhance reasoning capabilities.
Furthermore, \system can be customized to meet the unique needs of different organizations, underscoring its adaptability. Through extensive evaluations, we demonstrate that \system consistently outperforms industry-leading LLMs and state-of-the-art cybersecurity agents. By detailing its design, development, and evaluation, this work provides a blueprint for leveraging LLMs to address complex, data-intensive cybersecurity challenges.
\end{abstract}




\received{20 February 2007}
\received[revised]{12 March 2009}
\received[accepted]{5 June 2009}

\maketitle

\section{Introduction}

We are living in an era of rapid digital transformation, where technological advancements are closely linked to the growing prevalence of cyber threats. In recent years, the landscape of cyber threats has evolved dramatically, marked by a 25\% annual increase in reported Common Vulnerabilities and Exposures (CVEs) \cite{ibm_cve_proliferation}. The increase in CVEs can be attributed to several factors, especially the growing complexity of IT systems \cite{qualys_cve_surge_2024}, the widespread adoption of open-source software \cite{dam2023towards,darkreading_cves_plan}, and the accelerated pace of development cycles \cite{ox_modernizing_appsec}. Together, these factors facilitate the exploitation of cyber threats in modern software ecosystems.

\begin{figure}[!t]
    \centering
    \includegraphics[width =80mm]{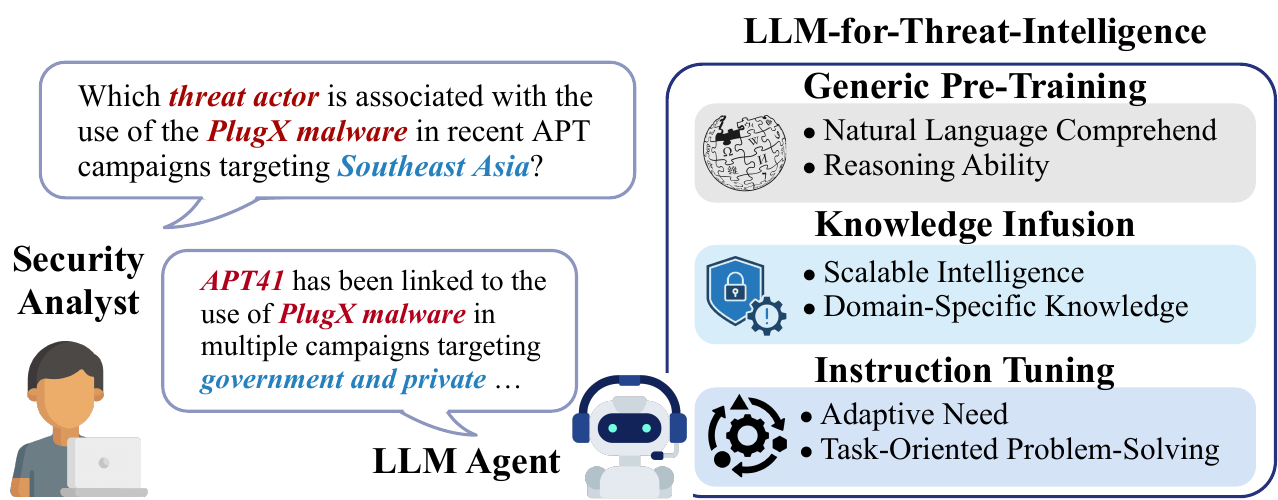}
    \caption{ (Left) An agentic LLM execution on threat intelligence tasks. (Right) The customization workflow to enhance the CTI proficiency of pre-trained LLMs.}
    \label{fig:example}
    \Description{Introduction Example.}
\end{figure}

To efficiently analyze and counteract these threats, {\it Cyber Threat Intelligence} (CTI) \cite{dehghantanha2018cyber} has emerged as a crucial strategy. CTI provides actionable insights derived from the analysis of threat data and empowers security teams to anticipate, identify, and mitigate cyber risks effectively \cite{sun2023cyber,bromiley2016threat}. However, it faces two critical challenges: \textbf{(1) Scalability}: The overwhelming volume of threat data creates significant bottlenecks \cite{abu2018cyber}. Security teams are often inefficiently processing and correlating this massive influx of security knowledge, limiting their ability to respond to emerging cyber threats. \textbf{(2) Adaptability}: Integrating CTI into existing cybersecurity platforms often requires extensive customization, which poses a significant barrier for small and medium-sized organizations who lack resources to deploy CTI effectively. To address these challenges, it is essential to propose advanced CTI solutions that adaptively support customization tailored to diverse organizational needs.

\begin{figure*}[!t]
    \centering
    \includegraphics[width =170mm]{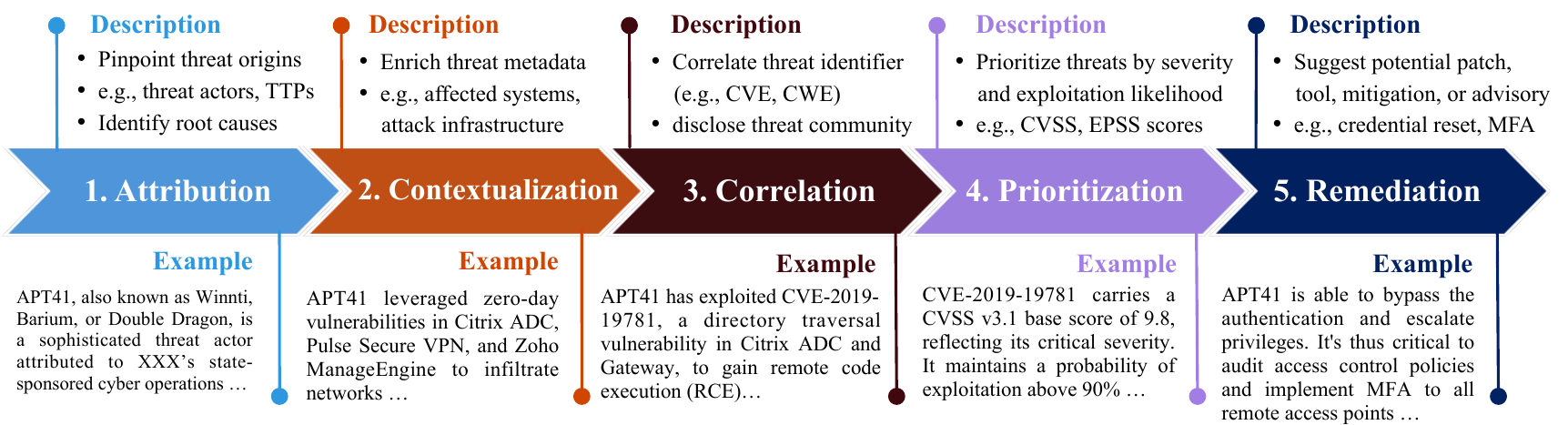}
    \caption{The lifecycle of threat intelligence tasks and running examples.}
    \label{fig:task}
    \Description{CTI pipeline and example.}
    \vspace{-4mm}
\end{figure*}


Large language models (LLMs) have recently demonstrated exceptional capabilities in natural language understanding and reasoning. Compared to conventional CTI solutions, such as rule-based systems \cite{ali2024ttpmapper, sun2023cyber} or function-frozen AI models \cite{montasari2021application, kumar2024ai, kant2023cyber, dutta2020overview}, LLMs offer superior scalability, as they can generalize across diverse knowledge-intensive domains without the need for task-specific architectures or manually crafted features \cite{budnikov2024generalization, kunter2013professional, brown2020language}. LLMs also exhibit strong adaptability. Techniques like instruction tuning and few-shot learning enable them to respond effectively to dynamic and data-scarce environments, where traditional models often struggle to keep pace \cite{liu2024enhancing,paeedeh2024cross, wang2021fast, al2017continuous}.

However, despite recent studies showing that industry-leading LLMs are not yet fully qualified for CTI analysis \cite{srikanth2024evaluating, kouremetis2025occult, xu2024large, shafee2024evaluation}, the blueprint for systematically customizing LLMs to support scalable and adaptive CTI analysis remains largely unexplored. This leads us to a central question:  {\textbf{How can LLMs be systematically developed to support diverse CTI activities?}}

\noindent
{\bf Our work.} To address this question, we propose \system, an LLM-powered agentic system designed to enhance CTI. \system is available in multiple model sizes—1B, 8B, and 70B—to accommodate varying computational constraints. As illustrated in Figure \ref{fig:example}, \system follows the {\bf Agentic LLM} paradigm, inheriting pre-trained language understanding and reasoning capabilities, and undergoing a series of targeted customizations to improve domain-specific expertise. We characterize \system by the following aspects:

\underline{\bf Lifecycle-Oriented Threat Analysis} --
CTI practice encompasses a lifecycle of analyzing and counteracting cyber threats \cite{irshad2023cyber, xiang2024ipattributor, qiang2017framework, sun2023cyber, patil2022threat}. As illustrated in Figure \ref{fig:task}, \system is aligned with this lifecycle and supports five groups of analytical tasks. To investigate the nature and origin of threats, \system enables \ding{182} {\it attribution} by identifying threat actors and collecting technical evidence, such as TTPs and campaign, to trace the source of an attack. It further facilitates \ding{183} {\it contextualization} by enriching metadata, including affected systems and infrastructures, to provide a comprehensive understanding of the threat landscape. Next, \ding{184} {\it correlation} involves linking known identifiers, such as CVEs \cite{mitre-cve}, while also uncovering adversarial networks by identifying related or concurrent threats. \system supports \ding{185} {\it prioritization} by assessing the severity of threats and estimating the likelihood of exploitation, enabling organizations to allocate resources efficiently. Finally, based on these analytical foundations, \ding{186} {\it remediation} supports the development of actionable countermeasures to effectively neutralize threats and reduce their broader impact.

\underline{\bf Scalable Knowledge Infusion} -- \system incorporates extensive CTI knowledge from a diverse set of threat databases and repositories, including MITRE-CVE \cite{mitre-cve}, CWE \cite{cwe}, CAPEC \cite{capec}, ATT\&CK \cite{mitre-attack}, NVD \cite{nvd}, Exploit-DB \cite{exploit-db}, and their correlated threat reporting platforms (e.g., IBM-XForce \cite{IBM_XForce}, Apache \cite{ apache_mailing_list}, and Red Hat \cite{redhat_bugzilla}), resulting in a collection of \ncve threat reports.

To integrate extensive CTI knowledge, \system adopts a \textbf{\textit{curriculum training}} methodology, where the ``curriculum'' consists of ordered threat reports designed to progressively embed CTI expertise into the model. This approach enables scalable knowledge infusion by allowing the training process to incorporate both extensive historical data and emerging threats. By adjusting pacing steps and balancing newer data with earlier content, curriculum training eliminates the need for retraining from scratch. 

\underline{\bf Adapting through Tuning} -- \system exhibits strong adaptability in addressing specific analytical needs, such as threat analysis related to particular products, vendors, or threat groups. To enable this, we apply instructional fine-tuning (or simply {\bf instruction tuning}), which leverages task-oriented datasets aligned with critical CTI tasks (\ding{182}–\ding{186}). Within these datasets, we introduce a {\bf cascading reasoning} framework that reflects the inherent task dependencies across the CTI lifecycle. This structured design allows \system to emulate best practices as expert-level CTI analysis.

\underline{\bf Modularized Inference} -- At inference time, we integrate specialized design modules to enhance \system's capacity for rigorous and reliable reasoning. For example, threat actor attribution involves applying named entity recognition (NER) to threat reports and retrieving relevant, up-to-date metadata regarding recent exploitation activities. By systematically organizing \system’s inference process through these modules, we improve the consistency and precision of its analytical outputs.

\underline{\bf Validated Proficiency} -- In experiments, we collect real-world threat reports from platforms including Oracle Security Alerts \cite{oracle_security}, Red Hat Bugzilla \cite{redhat_bugzilla}, Adobe Security Bulletin \cite{adobe_security_bulletin}, and the Apache Mailing List \cite{apache_mailing_list}. We design two groups of experiments to evaluate the scalability and adaptability of \system. For scalability evaluation, we evaluate \system's effectiveness across a large volume of threats spanning CTI tasks \ding{182}–\ding{186}, considering multiple analytical targets under each task. For example, threat contextualization includes identifying affected systems, attack infrastructure, and potential impacts. For adaptability evaluation, we focus on specialized threats targeting specific products (e.g., Acrobat), vendors (e.g., Adobe), or threat categories (e.g., DDoS). These scenarios reflect the diverse and evolving needs of industrial organizations and research institutions \cite{saeed2023systematic,van2021shared,sakellariou2022reference}.

In both evaluations, we categorize threats into two groups: {\bf (i) Historical Threats}—previously recorded threats that security analysts may not be fully aware of; and {\bf (ii) Zero-Day Threats}—newly emerging threats that have not yet been documented. Across both categories, \system consistently outperforms state-of-the-art (SOTA) models and industry-leading LLMs (e.g., \textsc{ChatGPT} and \textsc{Gemini}), demonstrating its versatility in addressing a broad range of cybersecurity challenges and its effectiveness in supporting proactive threat hunting within a dynamic threat landscape.

\noindent
{\bf Contributions.} We summarize the following contributions:
\begin{mitemize}
    \item We propose \system, a scalable and adaptive agentic system for CTI. \system supports lifecycle-oriented threat analysis, encompassing tasks from attribution to remediation. To accommodate varying computational constraints, \system is implemented in three model sizes: 1B, 8B, and 70B. Through comprehensive knowledge infusion, followed by instruction tuning and modularized inference, \system is equipped with deep CTI expertise and consistently delivers rigorous, reasoning-driven analysis.
    \item Through extensive evaluations on historical and zero-day threats, we not only demonstrate \system's effectiveness, but also reveal the varying performance of SOTA and industry-leading LLMs, highlighting the inherent complexity of CTI analysis and difficulty of achieving consistent expertise.
    \item The design and development of \system serve as a blueprint for integrating agentic LLM into cybersecurity practice. This endeavor not only advances the capabilities of CTI systems but also provides insights that could drive the development of expert-level AI across other high-stakes domains.\footnote{Codes and datasets are organized at \url{https://anonymous.4open.science/r/security-agent-4BB7/}}
\end{mitemize}

\section{Background and Related Work}



\subsection{Cyber Threat Ingelligence (CTI)}

AI-for-Security has been a long-standing research area, with extensive studies covering a broad range of cybersecurity topics, including anomaly detection \cite{yaseen2023role,pang2021deep}, malware classification\cite{kalash2018malware,pascanu2015malware}, and intrusion detection\cite{khraisat2019survey,frank1994artificial}. Among these, CTI has emerged as a systematic approach in cybersecurity, enabling organizations to collect, analyze, and operationalize cyber threats \cite{cisa2024,wagner2019cyber,sun2023cyber}.

Although there is no universal taxonomy in CTI, we categorize its involved tasks into five critical aspects based on extensive literature \cite{gao2020hincti,lin2023correlation,irshad2023cyber,qiang2017framework,sun2023cyber,patil2022threat,zhu2024nip,jacobs2020improving,kouremetis2025occult} and the MITRE initiative \cite{al2024mitre}, a foundational effort in the field. As illustrated in Figure \ref{fig:task}, the lifecycle of threat intelligence begins with investigatig threat properties (\ding{182} \& \ding{183}) \cite{irshad2023cyber,xiang2024ipattributor,qiang2017framework}, progresses to identify adversary community (\ding{184}) \cite{gao2020hincti,lin2023correlation,zhu2024nip}, and culminates in developing countermeasures (\ding{185} \& \ding{186}) \cite{sun2023cyber,patil2022threat,jacobs2020improving}. Formally, this work set the scope of the CTI tasks as follows:
\begin{mitemize}
    \item[\ding{182}] \textbf{Attribution}--Identifying the origin of cyber threats by profiling threat actors, analyzing tactics, techniques, and procedures (TTPs), and tracing attack campaign to link cyber incidents to specific adversaries or groups. Attribution helps organizations understand attacker intent, capabilities, and historical attack patterns \cite{xiang2024ipattributor,irshad2023cyber}.
    \item[\ding{183}] \textbf{Contextualization}--Enriching relevant metadata, such as affected systems, attack infrastructures, and potential impacts. Contextualization enables security analysts to assess threats with a comprehensive picture of the threat landscape\cite{li2022attackg,qiang2017framework}.
    \item[\ding{184}] \textbf{Correlation}--Linking available threat identifiers, such as CVE \cite{mitre-attack} and CWE \cite{cwe}, and uncovering related or concurrent activities within the same attack campaigns. Correlation supports to identify both known and emerging threats \cite{zhu2024nip, gao2020hincti}, while exposing the broader landscape of the attack community \cite{ lin2023correlation}.
    \item[\ding{185}] \textbf{Prioritization}-- Ranking threats based on severity and likelihood of exploitation to determine their remediation urgency. Effective prioritization optimizes resource allocation by focusing on the most critical and high-risk threats.\cite{sun2023cyber,jacobs2020improving}.
    \item[\ding{186}] \textbf{Remediation}--Developing and recommending actionable countermeasures, including security patches, mitigation strategies, and responsible advisories. Remediation ensures that organizations can effectively neutralize threats and reinforce their security posture \cite{jacobs2020improving,poehlmann2021organizational}.
\end{mitemize}
Within each CTI task, we define distinct analytical targets. For instance, prioritization involves assessing both severity, measured by the Common Vulnerability Scoring System (CVSS) \cite{NVD_CVSSv3_Calculator}, and exploitation potential, measured by the Exploit Prediction Scoring System (EPSS) \cite{EPSS2025}. The detailed analytical targets for each CTI task are presented in Table \ref{tab:ft-data} and are evaluated extensively in \cref{sec:expt}.




\subsection{Agentic LLM}
\label{ssec:agent-paradigm}


As extensively studied in recent works, the {\it agentic LLM paradigm} follows a two-stage development process: \(\theta_{\texttt{pt}} \rightarrow \theta_{\texttt{agent}}\) \cite{xu2023lemur,xi2023rise,yang2024finrobot}, where a pre-trained LLM \(\theta_{\texttt{pt}}\) serves as the foundation due to its inherent language comprehension and reasoning abilities. The model is then customized into \(\theta_{\texttt{agent}}\) to enhance its domain-specific expertise. This customization may involve re-training to infuse large-scale domain knowledge into LLMs—such as in programming synthesis tasks \cite{xu2023lemur}—or instruction tuning to adapt LLMs to highly specialized applications, such as cardiological diagnostics \cite{zhou2024zodiac}.  

During inference, agentic LLMs can be further enhanced using specialized prompts to structure to regularize their reasoning processes \cite{wei2022chain,hu2024improving}, or by integrating tool functions (e.g., APIs) to augment their problem-solving capabilities in domain-specific tasks \cite{qin2023tool,qin2023toolllm}. For example, given a threat report \(d\), one can implement a named entity recognition function \(\texttt{NER}(\cdot)\) callable by the LLM, such that \(\texttt{NER}(d) \rightarrow \mathcal{E}\) extracts threat-related entities (evidence) \(\mathcal{E}\) from \(d\). These tool functions may be implemented as embodied interfaces triggered through prompt instructions \cite{schick2023toolformer,li2024embodied}.  

\section{Methodology}
\label{sec:pt}

\subsection{Overview}

This section details the design of \system. Following the agentic LLM paradigm (\cref{ssec:agent-paradigm}), \system requires both customization and tailored inference to address heterogeneous CTI tasks.

Given the diverse analytical requirements and task-specific disparities in CTI, we adopt a two-stage customization consisting of {\bf re-training} and {\bf instruction tuning} to enhance LLMs with CTI expertise.  In the re-training stage, we build a large-scale threat corpus (\cref{ssec:pt-data}) and develop a {\bf curriculum training} approach (\cref{ssec:pt}) to progressively infuse CTI knowledge into LLMs. Next, we apply instruction tuning using a dataset curated with a cascading reasoning layout (\cref{ssec:ft}), which reflects the dependencies among CTI tasks. A fine-tuned LLM can then be guided to follow this reasoning workflow, enabling it to perform multi-step threat analyses in a structured and expert-like manner.

During inference, \system uses six NLP modules to structure its generation workflow (\cref{ssec:inference}). These include topic modeling to determine the query intent, entity and relation extraction, information retrieval, and finally, reasoning and evidence summarization to generate coherent, task-specific responses.
Together, these modules help \system produce accurate, structured outputs while minimizing hallucinations during open-minded generation.

\subsection{Threat Corups Construction}
\label{ssec:pt-data}

{\bf Raw Data Collection.} Given that threat data is predominantly represented by CVEs \cite{WikipediaCVE,WikipediaVulnerabilityDatabase}, we focus on collecting CVE-centered metadata from several threat databases: MITRE-CVE \cite{mitre-cve}, NVD \cite{nvd}, Exploit-DB \cite{exploit-db}, CWE \cite{cwe}, CAPEC \cite{capec}, and ATT\&CK \cite{mitre-attack}.

To enrich the collected cyber threat knowledge, we further extract reports and evidence from third-party references such as Oracle, Red Hat Bugzilla,
Adobe Security Bulletin, and the Apache Mailing List, incorporating diverse cybersecurity activities and countermeasure strategies. 
Table \ref{tab:source} outlines the data sources and the corresponding metadata collected.

{\bf Corpus Generation.} Next, we curate the collected metadata to construct CVE-centric threat reports, which are then aggregated as threat corpus used for re-training. As outlined in Algorithm \ref{alg:pt-corpus}, we employ a combination of three approaches to ensure the generation of high-quality threat reports and diverse layouts:

\begin{mitemize}
    \item {\bf In-Context Learning:} We extract sample layouts and phrasing styles from threat reports published by Microsoft \cite{microsoft2024defense}, Cisco \cite{talos2024}, CISA \cite{cisa2024}, Europol \cite{europol2017iocta}, and ENISA \cite{enisa2024}. These examples (or        ``demos'') $\mathcal{S}$ are used to guide the generation. For each generation, we select an example report \( s \in \mathcal{S} \) and construct a 1-shot prompt, designed to replicate the specific style and layout of \( s \).
    
    \item {\bf Multi-Agent Generation:} For each report generation, we randomly select one LLM from a model set $\Theta$, including GPT-4o, GPT-o1, Gemini-Pro, Pixtral-Large, DeepSeek-R1, or Llama-3.1-405B, to generate threat reports using the collected metadata and a 1-shot prompt. Subsequently, a different LLM is randomly chosen from $\Theta$ to revise the generated reports, introducing variability and enhancing quality by reducing single-model bias. 

    \item {\bf Output Sampling:} We adjust two decoding hyperparameters, temperature and top-p, which introduce controlled variability in the outputs, thus helps balance randomness and coherence, ensuring diverse yet meaningful generations \cite{chen2021evaluating, pearce2023examining, zhu2024hot}.
\end{mitemize}

\begin{algorithm}[t]
  \SetAlgoLined
  \KwIn{
        $\Theta$ -- a set of LLMs;
        $\mathcal{M}$ -- collected metadata; $\mathcal{S}$ -- example threat reports; $\mathcal{P}_\texttt{gen}$ --  generation prompt; $\mathcal{P}_\texttt{rev}$ -- revision prompt;
    }
    \KwOut{
        $D$ -- cyber threat corpus;
    }
    $D \leftarrow \emptyset$ \;
   \ForEach{metadata for a CVE: $m \in \mathcal{M}$}{
        \tcp{Prompt Construction}
        1-shot demo: $s \leftarrow \texttt{RandSelectFrom}(S)$ \;
        Input prompt: $\mathcal{P}_m \leftarrow \mathcal{P}_\texttt{gen} \oplus s \oplus m$\;
        \tcp{Report Generation and Revision}
        $\theta_1, \theta_2 \leftarrow \texttt{RandSelectFrom}(\Theta)$ \;
        Randomly set temperature $\mathcal{T}_1, \mathcal{T}_2$ and top-p $p_1, p_2$\;
        
        Generation: $d \leftarrow \texttt{OutSampling}(\theta_1, \mathcal{P}_m, \mathcal{T}_1, p_1)$ \;
        Input prompt: $\mathcal{P}_{m}\leftarrow \mathcal{P}_\texttt{rev} \oplus d$ \;
        Revision:  $d \leftarrow \texttt{OutSampling}(\theta_2, \mathcal{P}_{m}, \mathcal{T}_2, p_2)$ \;

        $D \leftarrow D \cup \{d\}$\;
   }
    \Return $D$\;
  \caption{Corpus Generation \label{alg:pt-corpus}}
\end{algorithm}

\begin{table}[!t]
  \small
  \def\arraystretch{0.9}
  \setlength{\tabcolsep}{4.5pt}
  \caption{Base LLMs and re-training settings. ``Cp'' -- Corpus; ``Tk'' -- token. We assign different data volumes each  model given their capacities and optimization requirements.}
  \begin{center}
  \scalebox{1.0}{
  \begin{tabular}{cccccccc}
    \toprule
    \multirow{2}{*}{\bf \system} & \multirow{2}{*}{\bf Base LLM} & \multicolumn{6}{c}{\bf Curriculum Training} \\
    \cmidrule(lr){3-8}
    & & \# Cp & \# Tk & $T$ & $T_1$ & $T_2$ & $\beta$ \\
    \midrule
    \system-1B & Llama-3.2-1B & 28K & 0.27B & 4 & 3 & 3 & 1.0 \\
    \system-8B & Llama-3.1-8B & 115K & 1.1B & 10 & 5 & 5 & 1.0 \\
    \system-70B & Llama-3.3-70B & 271K & 2.5B & 20 & 10 & 10 & 1.0 \\
    \bottomrule
  \end{tabular}}
  \end{center}
\label{tab:stat}
\end{table}

Due to space constraints, we defer the example reports, input prompts, and the corpus scale to Appendix \ref{app:pt-gen-detail}.


\begin{table*}[!t]
  \small
  \def\arraystretch{0.9}
  \setlength{\tabcolsep}{5pt}
  \caption{Details of CTI tasks. Each task targets a specific aspect of CTI analysis, incorporating varying dependencies from prior tasks, and involving different NLP modules: TOM--Topic Modeling; NER--Named Entity Recognition; REL--Relation Extraction; RAG--Retrieval-Augmented Generation (using web resources); REA--(any type of) Reasoning; SUM--Summarization. NLP modules are integrated at inference time to guide \system’s behavior.}
  \begin{center}
  \begin{tabular}{lllcccccc}
    \toprule
    \multirow{2.5}{*}{\bf CTI Task} & \multirow{2.5}{*}{\bf Analytical Target} & \multirow{2.5}{*}{\bf Dependency} & \multicolumn{6}{c}{\bf NLP Module} \\
    \cmidrule(lr){4-9}
    & & & TOM & NER & REL & RAG & REA & SUM \\ 
    \midrule
    \ding{182} Attribution & Threat Actor, TTPs, Campaign & Initial Evidence & \cmark & \cmark  & \cmark  & \cmark  & \cmark & \cmark \\
    \ding{183} Contextualization & Affected System, Attack Infra, Impact & Initial Evidence & \cmark & \cmark  & \cmark  &  & \cmark & \cmark \\
    \ding{184} Correlation & CVE, CWE Identifier & Initial Evidence+\ding{182}\ding{183} & \cmark  & &  & \cmark & \cmark & \cmark \\
    \ding{185} Prioritization & Severity (CVSS), Exploitation (EPSS) & Initial Evidence+\ding{182}\ding{183}\ding{184} & \cmark & \cmark & & \cmark  & \cmark & \cmark \\
    \ding{186} Remediation & Patch Tool, Code, Mitigation, Advisory & Initial Evidence+\ding{182}\ding{183}\ding{184}\ding{185} & \cmark & \cmark  & \cmark & \cmark & \cmark  & \cmark \\
    \bottomrule
  \end{tabular}
  \end{center}
\label{tab:ft-data}
\end{table*}

\subsection{Curriculum Training (Knowledge Infusion)}
\label{ssec:pt}


Next, we re-train base LLMs (detailed in Table \ref{tab:stat}) to infuse large-scale CTI knowledge. We design a ``curriculum'' strategy that progressively introduces the corpus throughout the training process.

{\bf Curriculum Design.} The ``curriculum'' refers to the ordered training corpus, structured by the time of threat report publication and the number of tokens in each report. Formally, let \(D = \{d_i\}_{i=1}^n\) represent the corpus, where \(d_i\) denotes individual threat reports. We first arrange \(D\) based on the year and month each CVE was reported. Within each year-month grouping, the reports \(\{d_i\}\) are further sorted by token length, creating a sequence that progresses chronologically from older to newer reports and by complexity from shorter to longer documents. For simplicity, we continue to denote the ordered corpus as \(D\).

{\bf Training with Pacing.} To enhance convergence and stabilize training, we introduce a pacing function 
\textsc{Pace}$(\cdot, \cdot)$ that governs the portion of the training corpus $D$ utilized during the $t$-th epoch. Formally:
\begin{equation}
\textsc{Pace}(D, t) = \begin{cases}
D[: |D| \cdot \frac{t}{T_1}] & t \leq T_1 \\
D & T_1 < t \leq T_2 \\
D \cup D[\beta \cdot |D|\cdot \frac{t - T_2}{T - T_2} : ] & t > T_2 
\end{cases}
\label{eq:pace}
\end{equation}

The pacing function in Eq. \ref{eq:pace} has three stages:  
{\bf (a) Linear Start} – In the early epochs (\(t \leq T_1\)), a progressively increasing portion of \(D\) is selected for training.  
{\bf (b) Plateau} – In the mid stage (\(T_1 < t \leq T_2\)), as the LLM \(\theta\) acquires foundational cyber threat knowledge, the entire corpus \(D\) is typically used for training.  
{\bf (c) Reinforced End} – In the final epochs (\(t > T_2\)), the corpus \(D\) is augmented with the most recent threat data to reinforce training of emerging cyber threats. This augmentation is balanced by a factor \(\beta\).  Table \ref{tab:stat} shows values of \(T\), \(T_1\), \(T_2\), and \(\beta\).


{\bf Training Objective.} Identical to the re-training objective of Llama \cite{touvron2023llama}, we employ Causal Language Modeling (CLM), an autoregressive training objective where the LLM predicts the next token in a sequence based on preceding tokens. Let \( x_j \) denote a token in a threat report \( d \) with \( L_d \) tokens, we have \( d = \{x_j\}_{j=1}^{L_d} \). The CLM objective is formulated as:  
\begin{equation}
\ell_\texttt{CLM}(\theta, D) = - \mathbb{E}_{d \sim \mathcal{D}} \left[ \sum_{j=1}^{L_d} \log p_\theta(x_j | x_{<j}) \right] + \lambda \|\theta - \theta_0\|_2^2
\end{equation}
Here, the model \(\theta\) is trained to maximize the likelihood of each token \(x_j\) given its preceding context \(x_{<j}\), across all tokens in a threat report \(d\). To prevent overfitting to the training corpus, we introduce a regularization term \(\|\theta - \theta_0\|_2^2\), where \(\theta_0\) denotes the original pre-trained parameters. This regularization encourages the model to retain its inherent language understanding and reasoning capabilities. The resulting objective function strikes a balance between effective CTI knowledge infusion and preserving the general-purpose strengths of the base LLM.

\subsection{Instruction Tuning via Cascading Reasoning}
\label{ssec:ft}

{\bf Data Construction.} CTI analysis follows a {\bf cascading} process, which starts with an initial set of evidence and progressively gathers additional information to refine the analysis \cite{oosthoek2021cyber,saeed2023systematic}. To equip LLMs with cascading analytical capabilities, we design a dataset to reflect progressive CTI activities: \(D = \{ E_0 \cup (Q_i, A_i)_{i=1}^n \}\) where \(E_0\) represents the initial evidence, and each query-answer pair \((Q_i, A_i)\) corresponds to a step in the cascading reasoning process: 
\[
(A_1|E_0, Q_1) \rightarrow (A_2 | E_0, A_1, Q_2) \rightarrow ... \rightarrow (A_n | E_0, A_1, ..., A_{n-1}, Q_n)
\]



{\setlength{\parindent}{0pt}
\begin{example}
Below is a cascading reasoning example.\\
\fbox{\parbox{\dimexpr\linewidth-2\fboxsep-2\fboxrule\relax}{
\noindent\strut\label{example:cascading}
\( \mathbf{E_0} \): A security monitoring system detected suspicious outbound traffic from several workstations within the organization’s network. Traffic analysis indicates command-and-control (C2) communication with IP addresses known to be associated with APT28, with observed indicators of $\cdots$ \\
{\bf Step 1: Threat Actor Attribution} \\
\(\mathbf{Q_1}\): From above, what threat actor is likely responsible? \\
\(\mathbf{A_1}\): The observed indicators —including C2 domain, malware characteristics, and past attack patterns— suggest that the activity is associated to APT28, a {\color{black!40}{\shadowtext{XXXX}}} -sponsored threat group known for cyber espionage. \\
{\bf Step 2: Vulnerability Identification} \\
\(\mathbf{Q_2}\): What known vulnerabilities (CVEs) are commonly associated with APT28 in recent campaigns? \\
\(\mathbf{A_2}\): APT28 has exploited CVE-2023-12345 (Windows Kernel Privilege Escalation) and CVE-2023-67890 (Remote Code Execution via malicious DLL sideloading). \\
{\bf Step 3: Prioritization of Threat Response} 
$\cdots$
}}
\end{example}}

\begin{figure*}[!t]
    \centering
    \includegraphics[width=\linewidth]{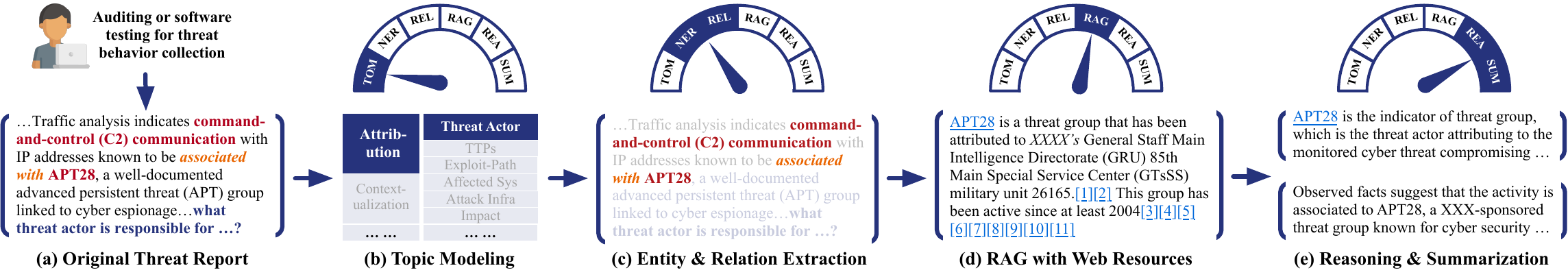}
    \caption{An illustration of modularization in \system, using the same threat report as described in Example \ref{example:cascading}.}
    \label{fig:nlp-module}
    \Description{Modularized inference.}
\end{figure*}

Table \ref{tab:ft-data} presents the analytical targets for CTI tasks \ding{182}-\ding{186}, where each task builds upon the preceding ones. As illustrated in Example \ref{example:cascading}, identifying CVEs requires first pinpointing threat actors, which subsequently guides the prioritization.

To construct the dataset, we aggregate references from threat databases (e.g., MITRE and NVD). For each reference, we extract available evidence as listed in Table \ref{tab:ft-data} and utilize descriptive reports as the initial evidence set \(E_0\). We then generate \((Q_i, A_i)\) pairs following the format exemplified in Example \ref{example:cascading}. It is important to note that not all constructed data samples encompass the full pipeline of tasks \ding{182}-\ding{186}. Dataset statistics are provided in Appendix \ref{app:ft}. 

{\bf Instruction Tuning.} To enable \system with cascading reasoning capabilities, we fine-tune it using the dataset \(D\) described above. This process follows a standard instruction-tuning objective \cite{xu2023lemur,ji2024sevenllm}, where the model is trained to generate a sequence of answers conditioned on the preceding context. Let \(\theta\) represent the parameters of \system. The objective function is defined as:  
\begin{equation}
\begin{split}
  \mathcal{L}(\theta) = - \mathbb{E}_{(E_0, Q_{1\sim n}, A_{1\sim n}) \in D}\ell(E_0, Q_{1\sim n}, A_{1\sim n}, \theta) \\ \text{s.t.}\quad \ell =\sum_{i=1}^{n} \log P_\theta(A_i | E_0, A_1, \dots, A_{i-1}, Q_i)  
\end{split}
\end{equation}
where \( P_\theta(A_i | E_0, A_1, \dots, A_{i-1}, Q_i) \) represents the probability assigned by the model to the correct response \(A_i\) given the accumulated context up to step \(i\).

\subsection{Modularized Inference}
\label{ssec:inference}

\system's inference also follows the cascading reasoning layout. Each reasoning step \((E_0, A_1, \dots, A_{i-1}, Q_i) \rightarrow A_i\) employs various NLP functions to modularize the analytical process. Table \ref{tab:ft-data} summarizes the NLP techniques applied to each CTI task. These techniques are implemented through the following approaches:

{\bf Topic Modeling.} We implement topic modeling to interpret users' intentions regarding the tasks they aim to perform. Specifically, we use \texttt{BERTopic}, a transformer-based package that eliminates the need for manually setting hyperparameters, as required in conventional methods \cite{blei2003latent, lee2000algorithms}. Given a threat report \(E_0\), we implement a two-step approach: first, based on CTI tasks \ding{182}-\ding{186}, and then, based on specific analytical targets, as outlined in Table \ref{tab:ft-data}.

{\setlength{\parindent}{0pt}\begin{example} Topic modeling (Figure \ref{fig:nlp-module}-(b)).\\
\fbox{\parbox{\dimexpr\linewidth-2\fboxsep-2\fboxrule\relax}{
Using the same example as in \ref{example:cascading}, the topic modeling step first determines the task as {\bf threat attribution} based on {\bf $Q_1$}. It then further specifies the analytical target as the {\bf threat actor} within the threat attribution task.}}
\end{example}}

{\bf Named Entity Recognition (NER) and Relation Extraction.} Next, we extract entities relevant to the identified topics while simultaneously identifying relationships among them. We employ the multi-agent generation as described in \cref{ssec:pt-data}, using few-shot prompting to instruct entity and relation extraction, where domain-specific entities (e.g., APT group) and their relationships are included as demonstrations within the prompt. We defer detailed implementation in \cref{app:ner-detail} and \ref{app:rel-detail}.

{\setlength{\parindent}{0pt}\begin{example}
NER and relation extraction (Figure \ref{fig:nlp-module}-(c)). \\
\fbox{\parbox{\dimexpr\linewidth-2\fboxsep-2\fboxrule\relax}{
Using Example \ref{example:cascading}, the NER identifies key entities {\bf "APT28"} and {\bf "C2 communication"} within the threat report. Simultaneously, relation extraction determines the {\bf "is associated with"} relationship between these entities. }}
\end{example}}

{\bf Retrieval Augmented Generation (RAG).} We implement RAG to retrieve relevant web documents through web screening to enrich threat reports. This is especially useful for zero-day threat hunting, where web-scale CTI intelligence—such as reported threat groups, affected products, and potential patches—is continuously updated. Then, we prompt GPT-o1 to rank the retrieved documents based on their relevance to the threat report $E_0$. To {\bf control latency}, we implement a cache to store retrieved documents. Caching documents are indexed by topic modeling results (i.e., CTI tasks and analytical targets) and specific entities (e.g., APT28 or C2 communication). 

{\bf Reasoning and Summarization.} We implement reasoning and summarization through prompting in \system, where reasoning is applied when deeper analysis or multi-step thinking is required. For example, reasoning is used to analyze severity level based on gathered evidence. Summarization is then applied to synthesize insights, tailoring the output to a specific CTI task by leveraging results from topic modeling, NER, relation extraction, and RAG.

Additional implementation details (e.g., prompts) of the six NLP modules are provided in \cref{app:nlp}.

{\bf Cascading Reasoning in Inference.} Since CTI tasks exhibit dependencies (Table \ref{tab:ft-data}), we minimize latency by concurrently executing topic modeling, NER, relation extraction, and RAG across all prerequisite tasks. Reasoning and summarization then proceed sequentially, following task dependencies to generate structured, step-by-step findings and conclusions.

\section{Experimental Setting}
\label{sec:expt-setting}

\begin{table*}[!t]
  \small
  \def\arraystretch{0.85}
  \setlength{\tabcolsep}{5pt}
  \caption{Evaluation results for \ding{182} {\bf Attribution} tasks, where "Hist."--historical threats, "0-Day"--zero-day threats. All metrics are scaled to 0-100 (\%) for consistency. The {\bf boldface} and \underline{underline} highlight the best and second-best results, respectively. }
  \begin{center}
  \scalebox{0.95}{
  \begin{tabular}{ccccccccccccccccc}
    \toprule
     & \multicolumn{4}{c}{\bf Threat Actor} & \multicolumn{8}{c}{\bf TTP} & \multicolumn{4}{c}{\bf Campaign} \\
    \cmidrule(lr){2-5} \cmidrule(lr){6-13} \cmidrule(lr){14-17}
     {\bf Model} & \multicolumn{2}{c}{Accuracy} & \multicolumn{2}{c}{BERT Score} & \multicolumn{2}{c}{Precision} & \multicolumn{2}{c}{Recall} & \multicolumn{2}{c}{F1 Score} & \multicolumn{2}{c}{IoU} & \multicolumn{2}{c}{Accuracy} & \multicolumn{2}{c}{BERT Score} \\
    \cmidrule(lr){2-3} \cmidrule(lr){4-5}\cmidrule(lr){6-7}\cmidrule(lr){8-9}\cmidrule(lr){10-11}\cmidrule(lr){12-13}\cmidrule(lr){14-15}\cmidrule(lr){16-17}
    & Hist. & \hlcell 0-Day & Hist. & \hlcell 0-Day & Hist. & \hlcell 0-Day & Hist. & \hlcell 0-Day & Hist. & \hlcell 0-Day & Hist. & \hlcell 0-Day & Hist. & \hlcell 0-Day & Hist. & \hlcell 0-Day \\
    \midrule
    \sevenllm & 65.94 & \hlcell 61.17 & 81.25 & \hlcell 79.83 & 73.44 & \hlcell 68.62 & 76.85 & \hlcell 71.53 & 75.11 & \hlcell 70.04 & 60.14 & \hlcell 53.90 & 73.56 & \hlcell 78.42 & 82.19 & \hlcell 80.65 \\
    \lily & 63.62 & \hlcell 54.58 & 78.25 & \hlcell 75.27 & 64.71 & \hlcell 55.32 & 70.17 & \hlcell 64.62 & 59.65 & \hlcell 67.28 & 42.50 & \hlcell 50.69 & 76.25 & \hlcell 70.38 & 82.16 & \hlcell 82.43 \\
    \midrule
    \gpt & 48.33 & \hlcell 40.31 & 76.02 & \hlcell 73.63 & 74.46 & \hlcell 77.42 & 81.53 & \hlcell 80.52 & 77.83 & \hlcell 78.94 & 63.71 & \hlcell 65.21 & 67.76 & \hlcell 69.74 & 76.15 & \hlcell 79.90 \\
    \gpto & 57.36 & \hlcell 55.34 & 77.93 & \hlcell 78.69 & 81.26 & \hlcell 72.49 & 83.18 & \hlcell 82.90 & 82.21 & \hlcell 77.35 & 69.79 & \hlcell 63.06 & 78.82 & \hlcell 72.78 & 77.22 & \hlcell 78.06 \\
    \gemini & 61.06 & \hlcell 57.43 & 81.24 & \hlcell 78.63 & 78.11 & \hlcell 79.36 & 81.19 & \hlcell 82.24 & 79.62 & \hlcell 80.77 & 66.14 & \hlcell 67.75 & 62.82 & \hlcell 69.21 & 75.84 & \hlcell 74.43 \\
    \llama & 34.85 & \hlcell 24.62 & 74.20 & \hlcell 71.35 & 72.46 & \hlcell 60.29 & 78.93 & \hlcell 71.08 & 75.56 & \hlcell 65.24 & 60.72 & \hlcell 48.41 & 41.64 & \hlcell 25.72 & 74.54 & \hlcell 71.58 \\
    \deepseek & 57.85 & \hlcell 55.26 & 80.39 & \hlcell 79.92 & 73.24 & \hlcell 71.51 & 76.43 & \hlcell 79.85 & 74.80 & \hlcell 75.45 & 59.75 & \hlcell 60.58 & 65.71 & \hlcell 63.04 & 78.27 & \hlcell 77.91 \\
    \claude & 30.12 & \hlcell 22.64 & 73.51 & \hlcell 72.96 & 69.82 & \hlcell 54.84 & 77.95 & \hlcell 75.26 & 73.66 & \hlcell 63.45 & 58.30 & \hlcell 46.46 & 58.91 & \hlcell 52.64 & 76.46 & \hlcell 72.83 \\
    \midrule
    \systems-1B & 82.81 & \hlcell 79.52 & 83.84 & \hlcell 81.72 & 86.15 & \hlcell 83.22 & 85.72 & \hlcell \underline{87.44} & 85.93 & \hlcell 85.28 & 75.34 & \hlcell 74.33 & 85.28 & \hlcell 80.61 & 84.62 & \hlcell 81.55 \\
    \systems-8B & \underline{87.58} & \hlcell \underline{83.74} & \underline{86.72} & \hlcell {\bf 85.39} & {\bf 90.67} & \hlcell {\bf 88.70} & {\bf 93.75} & \hlcell {\bf 92.74} & {\bf 92.18} & \hlcell {\bf 90.68} & {\bf 85.50} & \hlcell {\bf 82.94} & {\bf 95.71} & \hlcell {\bf 87.79} & {\bf 88.45} & \hlcell \underline{85.21} \\
    \systems-70B & {\bf 96.53} & \hlcell {\bf 88.29} & {\bf 87.93} & \hlcell \underline{84.52} & \underline{89.63} & \hlcell \underline{87.57} & \underline{88.26} & \hlcell 85.46 & \underline{88.94} & \hlcell \underline{86.50} & \underline{80.08} & \hlcell \underline{76.21} & \underline{91.53} & \hlcell \underline{86.06} & \underline{87.32} & \hlcell {\bf 87.03} \\
    \bottomrule
  \end{tabular}
  }
  \end{center}
\label{tab:expt-1}
\end{table*}

This section outlines our experimental setup for CTI tasks \ding{182}–\ding{186}, designed to evaluate \system's scalability and adaptability. The corresponding results are presented in \cref{sec:expt} and \ref{sec:expt-adapt}, respectively. 

\begin{tcolorbox}[
    colback=\boxcolor, 
    colframe=black, 
    width=0.48\textwidth, 
    boxrule=1pt, 
    arc=5pt,
    boxsep=0pt,
    left=5pt,
    right=5pt,
    coltitle=black,
]
\textbf{\textit{SYSTEM}}
\textcolor{blue!80}{You are a cybersecurity threat analyst specializing in cyber threat intelligence (CTI). Your role is to analyze threat reports, {\bf <Specify Analytical Target>}, and produce structured threat intelligence analysis grounded in established frameworks (e.g., MITRE ATT\&CK). You will follow a structured analytical pipeline with the following steps:}
\begin{mitemize}
    \item \textcolor{blue!80}{{\bf Resolve Task Dependencies:} Determine the prerequisite CTI tasks and analytical targets based on the input query.}
    \item \textcolor{blue!80}{{\bf Launch NLP Modules:} {\bf <Explain Six NLP Modules>}.}
\end{mitemize}
    
\vspace{4pt}
\textbf{\textit{USER}} 
\textcolor{asparagus}{{\bf <Report Title>}  {\bf <Report Excerpt>}}  \textcolor{crimson}{{\bf <User's Question>}}

\vspace{5pt} 
\textbf{\textit{ASSISTANT}} \textcolor{olive}{The user's intent pertains to \textbf{<Topic Modeling>}, which aligns with the prerequisite identification of \textbf{<CTI Tasks>} and corresponding \textbf{<Analytical Targets>}.}

\textcolor{violet!80}{Proceeding with a structured inference pipeline, we will now extract and analyze the following components: \textbf{<Key Entities>}, \textbf{<Entity Relations>}, \textbf{<Retrieved Evidence>}, \textbf{<Sources>}, and \textbf{<Synthetic Reasoning>}.}

\textcolor{cyan}{\textbf{Final Response}: \textbf{<Summarization>}}
\end{tcolorbox}
\noindent\begin{minipage}{0.48\textwidth}
\vspace{-15pt}
\captionof{figure}{Inference Prompt structure. For baselines and \system, we prompt them to perform cascading reasoning with the NLP modules defined in Table \ref{tab:ft-data}. For tasks with prior dependencies, we instruct LLMs to first complete the prerequisite tasks before proceeding. }
\label{fig:prompt}
\end{minipage}

\vspace{7pt}
{\bf Scenarios.} To evaluate the scalability and adaptability of \system, we consider two experimental scenarios:

\begin{mitemize}
    \item {\bf Broad CTI} — Demonstrates \system's scalability in analyzing a wide range of cyber threats across CTI tasks \ding{182}–\ding{186}, providing a comprehensive view of the threat landscape.
    \item {\bf Specialized CTI} — Demonstrates \system's adaptability to specific cyber threat groups, such as those targeting particular products or vendors tailored to business needs, or threats within a specific category to support focused interests (e.g., \cite{mishra2022cyber, balasubramanian2025cognitive}).
\end{mitemize}

In both scenarios, \system undergoes the same re-training using the threat corpus described in Table \ref{tab:stat}. However, instruction tuning differs based on the evaluation focus. For the broad CTI setting, we use a diverse instruction tuning dataset covering a wide spectrum of threats, as reported in Table \ref{tab:ft-stat}. In the Specialized CTI setting, we extract targeted subsets of threats from the broader dataset and {\bf independently} fine-tune \system to emphasize its functions on specific threat groups (e.g., those associated with a particular product, vendor, or threat category).

{\bf Dataset.} To ensure practical evaluation, we directly collect threats reported from real-world platforms, including Oracle Security Alerts \cite{oracle_security}, Red Hat Bugzilla \cite{redhat_bugzilla}, Adobe Security Bulletin \cite{adobe_security_bulletin}, NVD \cite{nvd}, VulDB \cite{vuldb}, Patchstack \cite{patchstack}, and public mailing lists from Apache \cite{apache_mailing_list}, Debian \cite{debian_security}, openSUSE \cite{opensuse_security}, and Fedora \cite{fedora_security}.
To avoid data leakage, all threats correlated with these evaluation reports are excluded from both the re-training and instruction tuning phases. We categorize the evaluation threats into two groups: {\bf (1) Historical} — Threats disclosed between 2015 and 2024. {\bf (2) Zero-Day} — Newly disclosed threats from the two months immediately preceding our evaluation period in 2025. Table \ref{tab:eval-data-stat} summarizes the statistics of the evaluation dataset.

\begin{table}[h]
  \small
  \setlength{\tabcolsep}{4pt}
  \caption{Statistics of evaluation threats (not training) for broad and specialized CTI. "Mc."--MacOS, "Cr."--Chrome, "Lk."--Linux Kernel, "Ad."--Adobe, "Dl."--Dell, "Of."--Overflow}
  \vspace{-7pt}
  \begin{center}
  \scalebox{0.85}{
  \begin{tabular}{ccccccccccc}
    \toprule
     \multirow{2.5}{*}{\bf \# Threat} & \multirow{2.5}{*}{\bf Broad} & \multicolumn{3}{c}{\bf Product} & \multicolumn{3}{c}{\bf Vendor} & \multicolumn{3}{c}{\bf Category} \\
     \cmidrule(lr){3-5}\cmidrule(lr){6-8}\cmidrule(lr){9-11}
     & & Mc. & Cr. & Lk. & IBM \ & Ad. & Dl. & Of. & DoS & XSS \\
    \midrule
     Historical & 26.5K & 2.5K & 1.7K & 4.6K & 3.3K & 1.8K & 0.8K & 7.5K & 9.6K & 12.8K \\
     \hlcell Zero-Day & \hlcell 3.6K & \hlcell 141 & \hlcell 48 & \hlcell 176 & \hlcell 81 & \hlcell 72 & \hlcell 33 & \hlcell 407 & \hlcell 721 & \hlcell 2.6K  \\
    \bottomrule
  \end{tabular}}
  \end{center}
\label{tab:eval-data-stat}
\end{table}

{\bf Baseline.} We primarily consider two groups of baselines: (1) SOTA CTI agents, \textsc{SevenLLM} \cite{ji2024sevenllm} and \textsc{Lily-Cyber} \cite{LilyCybersecurity7B} and (2) Industry-leading general-purpose LLMs: \textsc{ChatGPT-4o/o1}, \textsc{Gemini-Pro}, \textsc{Llama-405B}, \textsc{DeepSeek-R1}, and \textsc{Claude-Opus}. 
Additionally, we investigate the impact of {\bf ablation baselines} in \cref{ssec:ablation}.

{\bf Metrics.} We use different evaluation metrics depending on the specific CTI task. The definitions, strategies, and rationale for using these metrics are provided in Appendix \ref{app:eval-metric}.

\section{Evaluation Results: Broad CTI}
\label{sec:expt}

\begin{table*}[!t]
  \small
  \def\arraystretch{0.88}
  \setlength{\tabcolsep}{4.5pt}
  \caption{Evaluation results for \ding{183} {\bf Contextualization} tasks, where "Hist."--historical threats, "0-Day"--zero-day threats. All metrics are scaled to 0-100 (\%) for consistency. The {\bf boldface} and \underline{underline} highlight the best and second-best results, respectively.}
  \begin{center}
  \scalebox{0.9}{
  \begin{tabular}{ccccccccccccccccccc}
     \toprule
     & \multicolumn{8}{c}{\bf Affected System (w/ Version, if available)} & \multicolumn{8}{c}{\bf Attack Infrastructure} & \multicolumn{2}{c}{\bf Impact} \\
    \cmidrule(lr){2-9} \cmidrule(lr){10-17} \cmidrule(lr){18-19}
    {\bf Model} & \multicolumn{2}{c}{Precision} & \multicolumn{2}{c}{Recall} & \multicolumn{2}{c}{F1 Score} & \multicolumn{2}{c}{IoU} & \multicolumn{2}{c}{Precision} & \multicolumn{2}{c}{Recall} & \multicolumn{2}{c}{F1 Score} & \multicolumn{2}{c}{IoU}  & \multicolumn{2}{c}{BERT Score}  \\
    \cmidrule(lr){2-3} \cmidrule(lr){4-5}\cmidrule(lr){6-7}\cmidrule(lr){8-9}\cmidrule(lr){10-11}\cmidrule(lr){12-13}\cmidrule(lr){14-15}\cmidrule(lr){16-17}\cmidrule(lr){18-19}
    & Hist. & \hlcell 0-Day & Hist. & \hlcell 0-Day & Hist. & \hlcell 0-Day & Hist. & \hlcell 0-Day & Hist. & \hlcell 0-Day & Hist. & \hlcell 0-Day & Hist. & \hlcell 0-Day & Hist. & \hlcell 0-Day & Hist. & \hlcell 0-Day \\
    \midrule
    \sevenllm & 24.76 & \hlcell 13.08 & 18.57 & \hlcell 6.53 & 21.22 & \hlcell 8.71 & 11.87 & \hlcell 4.55 & 43.17 & \hlcell 40.28 & 53.62 & \hlcell 45.92 & 47.83 & \hlcell 42.92 & 31.43 & \hlcell 27.32 & 75.92 & \hlcell 77.64 \\
    \lily & 27.73 & \hlcell 22.42 & 19.62 & \hlcell 14.57 & 22.9 8& \hlcell 17.66 & 12.98 & \hlcell 9.69 & 44.73 & \hlcell 35.16 & 56.82 & \hlcell 53.13 & 50.06 & \hlcell 42.32 & 33.38 & \hlcell 26.84 & 73.02 & \hlcell 76.81 \\
    \midrule
    \gpt & 53.76 & \hlcell 48.62 & 41.80 & \hlcell 36.09 & 47.03 & \hlcell 41.43 & 30.75 & \hlcell 26.13 & 54.51 & \hlcell 47.78 & 62.55 & \hlcell 57.22 & 58.25 & \hlcell 52.08 & 41.10 & \hlcell 35.20 & 76.52 & \hlcell 78.16 \\
    \gpto & 58.74 & \hlcell 55.49 & 58.61 & \hlcell 47.37 & 58.67 & \hlcell 51.11 & 41.52 & \hlcell 34.33 & 61.72 & \hlcell 50.33 & 65.23 & \hlcell 63.26 & 63.43 & \hlcell 56.06 & 46.44 & \hlcell 38.95 & 78.91 & \hlcell 80.24 \\
    \gemini & 35.62 & \hlcell 33.17 & 41.52 & \hlcell 34.50 & 38.34 & \hlcell 33.82 & 23.72 & \hlcell 20.35 & 42.35 & \hlcell 34.35 & 31.17 & \hlcell 22.62 & 35.91 & \hlcell 27.28 & 21.88 & \hlcell 15.79 & 74.85 & \hlcell 72.62 \\
    \llama & 11.46 & \hlcell 8.12 & 21.02 & \hlcell 10.74 & 14.83 & \hlcell 9.25 & 8.01 & \hlcell 4.85 & 31.29 & \hlcell 28.74 & 27.08 & \hlcell 20.35 & 29.03 & \hlcell 23.83 & 16.98 & \hlcell 13.53 & 72.24 & \hlcell 68.33\\
    \deepseek & 36.71 & \hlcell 31.88 & 54.63 & \hlcell 47.62 & 43.91 & \hlcell 38.19 & 28.13 & \hlcell 23.60 & 48.92 & \hlcell 33.96 & 40.92 & \hlcell 31.87 & 44.56 & \hlcell 32.88 & 28.67 & \hlcell 19.68 & 74.26 & \hlcell 72.41\\
    \claude & 31.26 & \hlcell 29.82 & 27.35 & \hlcell 22.41 & 29.17 & \hlcell 25.59 & 17.08 & \hlcell 14.67 & 44.65 & \hlcell 41.71 & 48.90 & \hlcell 46.48 & 46.68 & \hlcell 43.97 & 30.44 & \hlcell 28.18 & 76.32 & \hlcell 74.15 \\
    \midrule
    \systems-1B & 78.64 & \hlcell 74.45 & 81.24 & \hlcell 77.26 & 79.92 & \hlcell 75.83 & 66.55 & \hlcell 61.07 & 85.73 & \hlcell 82.24 & 87.62 & \hlcell 86.61 & 86.66 & \hlcell 84.37 & 76.47 & \hlcell 72.96 & 83.75 & \hlcell 85.26 \\
    \systems-8B & \underline{87.91} & \hlcell \underline{85.82} & \underline{91.86} & \hlcell \underline{88.12} & \underline{89.84} & \hlcell \underline{86.95} & \underline{81.56} & \hlcell \underline{76.92} & \underline{87.46} & \hlcell \underline{85.71} & \underline{92.83} & \hlcell \underline{88.17} & \underline{90.07} & \hlcell \underline{86.92} & \underline{81.93} & \hlcell \underline{76.87} &  {\bf 90.03} & \hlcell {\bf 88.31} \\
    \systems-70B & {\bf 94.54} & \hlcell {\bf 88.17} & {\bf 96.04} & \hlcell {\bf 93.75} & {\bf 95.28} & \hlcell {\bf 90.87} & {\bf 90.99} & \hlcell {\bf 83.28} & {\bf 96.37} & \hlcell {\bf 90.16} & {\bf 97.69} & \hlcell {\bf 93.18} & {\bf 97.03} & \hlcell {\bf 91.65} & {\bf 94.22} & \hlcell {\bf 84.58} & \underline{87.33} & \hlcell \underline{85.84} \\
    \bottomrule
  \end{tabular}
  }
  \end{center}
\label{tab:expt-2}
\end{table*}

\subsection{Threat Attribution}

We begin by attributing the origin of cyber threats with three targets: threat actor, TTP, and campaign, as outlined in Table \ref{tab:ft-data}. Our results yield several research findings (RF):


{\bf RF\(_1\): Specialized development is necessary.} As shown in Table \ref{tab:expt-1}, industry-leading LLMs, despite their broad applicability across diverse tasks, exhibit limited performance in CTI analysis. For instance, \textsc{ChatGPT-4o} has achieved only 40.31\% accuracy in threat actor attribution, significantly lower than \system-1B (79.52\%) and \system-8B (83.74\%) for the same task. This gap underscores the necessity of specialized development, as detailed in \cref{sec:pt}. Tailoring models specifically for CTI significantly reduces parameter requirements—\system-1B and 8B are far more lightweight compared to the 500B–1T parameters of \textsc{GPT-4o}. 


{\bf RF\(_2\): Large model size is not always beneficial.} Interestingly, scaling up model size does not always translate to improved CTI effectiveness. For instance, \system-8B achieves 93.74\% recall in TTP attribution, outperforming \system-70B (88.26\%), suggesting diminishing benefits. This indicates the importance of domain-specific optimizations instead of sheer model sizes.

{\bf RF\(_3\): Task-specific capability varies across baselines.} As shown in Table \ref{tab:expt-1}, baseline models exhibit varying levels of effectiveness across different CTI tasks. For instance, \textsc{GPT-o1} performs obviously worse in attributing threat actors compared to campaign. A deeper case study reveals that baseline LLMs struggle to accurately interpret CTI-specific terminology related to threat actors. For example:
\begin{tcolorbox}[
    colback=\boxcolor,  
    colframe=black, 
    width=0.48\textwidth,  
    boxrule=1pt, 
    arc=5pt,
    left=5pt,    
    right=5pt,   
    boxsep=0pt, 
    top=3pt,
    bottom=3pt,
]
\begin{case}
When analyzing Log4Shell (a known vulnerability) with Mirai (a botnet) as the threat actor, GPT-o1  describes the attacker as an ``outside attacker'', failing to precisely identify Mirai. This demonstrates a fundamental gap in attribution accuracy for complex cyber threats.
\end{case}
\end{tcolorbox}

On the other hand, \textsc{GPT-o1} achieves $>$78\% accuracy in analyzing attack campaign, as this task primarily involves summarizing existing evidence instead of complex reasoning. 
\begin{tcolorbox}[
    colback=\boxcolor,  
    colframe=black, 
    width=0.48\textwidth,  
    boxrule=1pt, 
    arc=5pt,
    left=5pt,    
    right=5pt,   
    boxsep=0pt, 
    top=3pt,
    bottom=3pt,
]
\begin{case}
In cases where specificity is less critical, a rough description—such as "JNDI lookup and remote code fetching"—sufficiently summarizes the campaign of Log4Shell. While a more precise description would be:
"The Log4j library contacts the attacker's malicious LDAP server, which then responds with a reference to a remote Java class file containing malicious code,"
the broader summary still effectively conveys the core mechanism.
\end{case}
\end{tcolorbox}
This indicates that baseline models perform better in tasks with higher generation flexibility but face challenges in tasks requiring precise CTI expertise and structured analysis.


\subsection{Threat Contextualization}

We continue evaluating \system in contextualizing affected systems, attack infrastructure (e.g., domains and servers), and impact.Our evaluations derive complementary research findings:

{\bf RF\(_4\): Larger LLMs tend to be wordy.} From Table \ref{tab:expt-2}, we observe that larger LLMs tend to achieve lower text similarity scores as measured by BERTScore. For example, \system-70B performs worse than \system-8B in generating threat impact descriptions on historical threats (87.33 vs. 90.03). Through detailed case studies, we find that larger LLMs are more likely to produce verbose responses:
\begin{tcolorbox}[
    colback=\boxcolor,  
    colframe=black, 
    width=0.48\textwidth,  
    boxrule=1pt, 
    arc=5pt,
    left=5pt,    
    right=5pt,   
    boxsep=0pt, 
    top=3pt,
    bottom=3pt,
]
\begin{case}
\system-8B summarizes an impact analysis concisely: "The March 2022 cyberattack by the Lazarus Group resulted in the theft of \$620 million from the Ronin Network, significantly impacting the cryptocurrency market and..." This closely aligns with the reference \cite{LazarusGroupWiki}, {\bf maintaining factual accuracy while remaining succinct.}

In contrast, \system-70B generates a more verbose response: "In March 2022, the notorious Lazarus Group orchestrated a sophisticated cyber heist, infiltrating the Ronin Network's defenses to abscond with a staggering \$620 million in cryptocurrency. This audacious act not only sent shockwaves through the digital currency markets, causing a temporary plunge in asset values, but also undermined investor confidence..."

The latter output is {\bf overly wordy and includes sentiment-heavy language} (e.g., "notorious," "sophisticated," "staggering," "audacious"), making it less aligned with the original reference.
\end{case}
\end{tcolorbox}

This finding aligns with RF\(_2\), reinforcing that larger models are not always more effective for CTI tasks. However, to balance overall performance across all CTI tasks, a systematic evaluation is necessary to determine the optimal model scale. 


{\bf RF\(_5\): Zero-day threat hunting is not always harder.} Both Table \ref{tab:expt-1} and \ref{tab:expt-2} reveal that analyzing zero-day threats does not always result in worse performance compared to historical threats. For instance, when assessing the impact of cyber threats, LLMs such as \system-1B and \textsc{SevenLLM} sometimes achieve slightly better relevance in their generated analyses. We attribute this to the nature of specific CTI tasks: For impact analysis, the contextualized understanding of previously known threats allows LLMs to provide more insightful assessments, particularly when certain zero-day threats resemble threats they have already encountered. However, for tasks such as identifying affected systems, attack infrastructures, or threat actors, zero-day threats often introduce novel and previously unseen products or malware, making analysis more challenging compared to historical threats. 


{\bf RF\(_6\): CTI agents should align with real-world practice.}
Despite the strong performance of SOTA agents (\textsc{SevenLLM} and \textsc{Lily-Cyber}) on certain tasks (e.g., analyzing TTPs), they struggle with others, such as attack infrastructure analysis. We attribute this to a misalignment between their training focus and CTI applications. For instance, \textsc{SevenLLM} is trained with a broad emphasis on NLP tasks (e.g., NER), while \textsc{Lily-Cyber} is developed using {\it hand-crafted cybersecurity and hacking-related data pairs} \cite{LilyCybersecurity7B}. However, these training paradigms often conflate TTPs and threat actors, rather than treating them as distinct analytical components. This lack of granularity can hinder precise CTI reasoning, particularly for tasks that require structured analysis of attack properties.


\begin{table*}[!t]
  \small
  \def\arraystretch{0.9}
  \setlength{\tabcolsep}{5pt}
  \caption{Evaluating {\bf \ding{186} Prioritization} performance using CVSS 3.x metrics \cite{NVD_CVSSv3_Calculator}: AV--Attack Vector, AC--Attack Complexity, PR--Privileges Required, UI--User Interaction, S--Scope, C--Confidentiality Impact, I--Integrity Impact, A--Availability Impact, Base--CVSS Base Severity. We report the accuracy of correctly categorizing each metric (e.g., Attack Vector: Network). The {\bf boldface} and \underline{underline} highlight the best and second-best results, respectively.}
  \begin{center}
  \scalebox{0.92}{
  \begin{tabular}{ccccccccccccccccccccc}
     \toprule
     \multirow{2.5}{*}{\bf Model} & \multicolumn{9}{c}{\bf Historical Threat} & \multicolumn{9}{c}{\hlcell {\bf Zero-Day Threat}}\\
    \cmidrule(lr){2-10}\cmidrule(lr){11-19}
    & AV & AC & PR & UI & S & C & I & A & Base &  \hlcell  AV & \hlcell  AC & \hlcell  PR & \hlcell  UI & \hlcell  S & \hlcell  C & \hlcell  I & \hlcell  A & \hlcell  Base \\
    \midrule
    SL & 22.51 & 82.64 & 35.04 & 54.21 & 56.87 & 30.14 & 21.62 & 54.67  & 32.75 &\hlcell 14.76 &\hlcell 76.85 &\hlcell 24.46 &\hlcell 32.27 &\hlcell 46.83 &\hlcell 33.70 & \hlcell35.62 & \hlcell 51.84 &\hlcell 27.58 \\
    LY & 54.75 & 65.83 & 22.12 & 76.91 & 53.56 &  43.17 & 37.85 & 42.73 & 43.56 &\hlcell 47.52 &\hlcell 64.31 &\hlcell 19.75 &\hlcell 59.17 &\hlcell 58.75 &\hlcell  37.28 &\hlcell 33.54 &\hlcell 45.19 &\hlcell 36.61  \\
    \midrule
    4O & 76.85 & 92.91 & 61.84 & 80.78 & 66.51 & 27.05 & 20.43 & 34.21 & 58.96 &\hlcell 77.81 &\hlcell 84.55 &\hlcell  56.78 &\hlcell  62.35 &\hlcell 68.14 &\hlcell 22.56 &\hlcell 31.05 &\hlcell 28.49 &\hlcell 46.51 \\
    O1 & 82.51 & 93.36 & 70.57 & 88.21 & 72.53 &  51.69 & 24.22 & 46.73 & 69.25 &\hlcell 86.83 &\hlcell 92.74 &\hlcell  63.17 &\hlcell  77.14 &\hlcell 76.83 &\hlcell 45.30 &\hlcell 47.26 &\hlcell  54.03 &\hlcell 63.71  \\
    GM & 82.07 & 90.71 & 46.42 & 73.72 & 61.24 & 44.75 & 39.80 & 38.19 & 52.51 &\hlcell 67.38 &\hlcell 83.26 &\hlcell 31.07 & \hlcell 56.34 &\hlcell 55.62 &\hlcell 31.27 &\hlcell 36.56 &\hlcell 42.93 &\hlcell 39.81  \\
    LA & 37.18 & 56.25 & 67.24 & 44.21 & 48.67 &  22.13 & 18.69 & 29.32 & 24.36 &\hlcell 12.61 &\hlcell 57.63 &\hlcell  54.53 &\hlcell 39.81 &\hlcell 28.75 &\hlcell 10.33 &\hlcell  17.72 &\hlcell  11.36 &\hlcell 17.45  \\
    DS & 64.52 & 87.21 & 54.52 & 87.36 & 69.54 &  46.27 & 44.18 & 40.21 & 62.64 &\hlcell 58.82 &\hlcell 82.01 &\hlcell 46.42 & \hlcell 82.86 &\hlcell 51.35 &\hlcell 35.26 &\hlcell 40.96 &\hlcell 38.74 &\hlcell 53.26 \\
    CD & 53.76 & 66.25 & 63.87 & 56.32 & 62.71 &  35.06 & 41.28 & 37.93 & 49.07 &\hlcell 52.16 &\hlcell 55.86 &\hlcell 66.46 & \hlcell 49.39 &\hlcell  57.24 &\hlcell 30.04 &\hlcell  22.75 &\hlcell  28.39 &\hlcell  36.88 \\
    \midrule
    C1 & 86.26 & 93.32 & 88.27 & 94.55 & 95.51 & 81.10 & 79.63 & 84.23 & 88.17 &\hlcell 84.71 &\hlcell 90.83 &\hlcell 86.06 & \hlcell 96.64 &\hlcell 90.32 &\hlcell 76.64 &\hlcell 73.59 &\hlcell 81.62 &\hlcell 83.44 \\
    C8 & {\bf 94.74} & \underline{98.85} & \underline{95.53} & {\bf 99.73} &\underline{97.63} & \underline{93.39} & \underline{90.33} & \underline{95.49} & \underline{96.86} &\hlcell {\bf 93.27} &\hlcell \underline{99.31} &\hlcell \underline{94.42} &\hlcell {\bf 97.65} &\hlcell {\bf 94.52} &\hlcell \underline{85.43} &\hlcell \underline{88.47} &\hlcell  \underline{86.54} &\hlcell \underline{91.57} \\
    C70 & \underline{94.33} & {\bf 100.0} & {\bf 97.56} & \underline{97.46} & {\bf 98.15} &  {\bf 94.74} & {\bf 92.26} & {\bf 95.83} & {\bf 97.57} &\hlcell \underline{92.69} &\hlcell {\bf 100.0} &\hlcell  {\bf 97.71} &\hlcell \underline{95.48} &\hlcell \underline{93.26} &\hlcell {\bf 90.97} &\hlcell {\bf 94.53} &\hlcell {\bf 92.66} &\hlcell {\bf 93.16} \\
    \bottomrule
  \end{tabular}
  }
  \end{center}
\label{tab:expt-5}
\end{table*}

\subsection{Threat Correlation}

We now evaluate \system on {\bf \ding{184} correlation} tasks. The objective is to measure how effectively \system can identify relevant threat identifiers (CVEs and CWEs) from a given threat report. Moreover, we extend the correlation by identifying additional threats (CVEs) that are not directly from threat reports but are related or concurrent, i.e., the {\bf Expansive Correlation}. We use different strategies:
(i) Directly conduct expansive correlation.
(ii) First find CVEs from threat reports and then expand to additional CVEs.
(iii) First find CWEs from threat reports  and then expand to additional CVEs.
Our evaluations yield the following findings: 
\begin{figure}[!t]
    \centering
    \includegraphics[width =84mm]{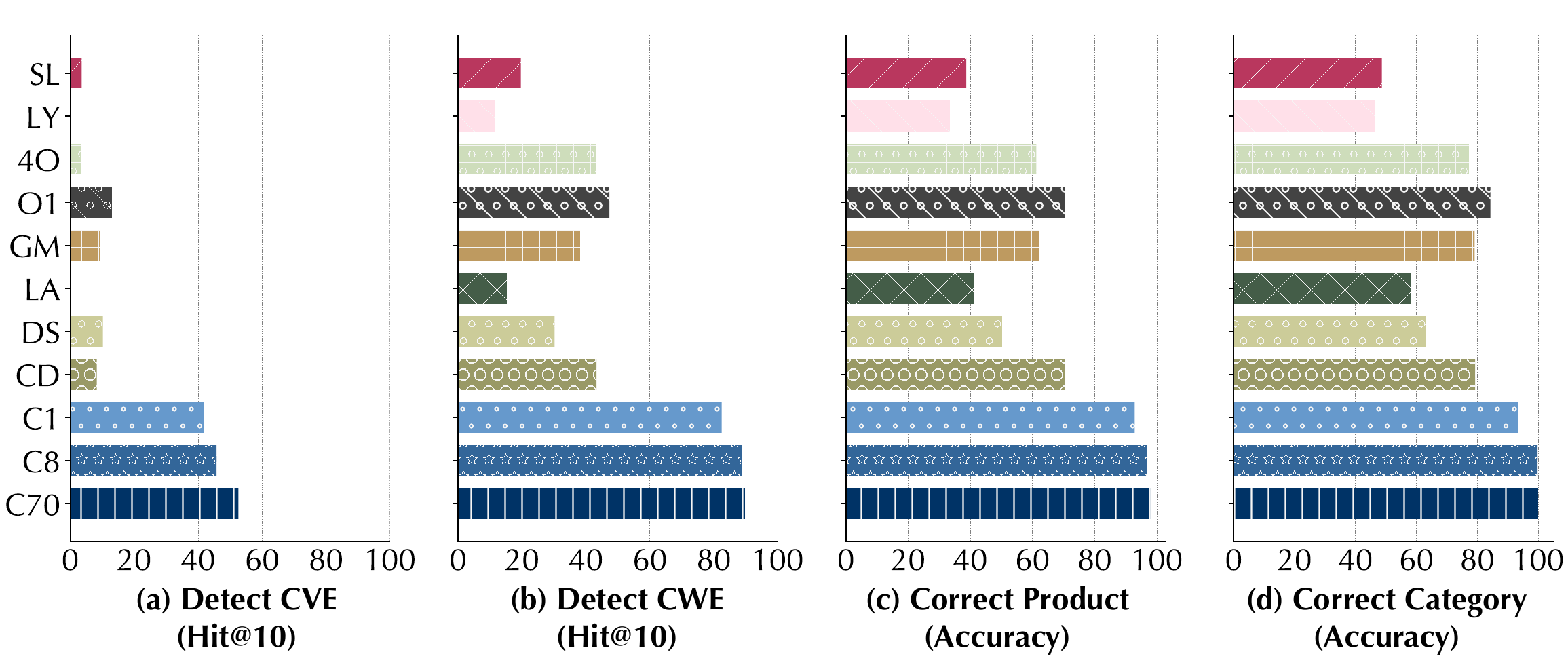}
    \caption{Evaluation results on \ding{184} Correlation using historical threats (we do not evaluate on 0-day threats as they have incomplete correlation reports). The presentation order of LLM names (top-down) follows the same presentation sequence as in Tables \ref{tab:expt-1} (top-down).}
    \label{fig:expt-3}
\end{figure}

{\bf RF\(_7\): \system demonstrates strong correlation relevance.}
Figure \ref{fig:expt-3}-(a)(b) shows the hit ratio for the top 10 correlated threat identifiers (CVEs, CWEs). Correlating CVEs is more challenging due to their vast number ($>$200k) compared to CWEs ($\sim$900). However, we also observe that \system is able to correlate more relevant CVEs such as those that belong to the same category or affect the same products. For example:

\begin{tcolorbox}[
    colback=\boxcolor,  
    colframe=black, 
    width=0.48\textwidth,  
    boxrule=1pt, 
    arc=5pt,
    left=5pt,    
    right=5pt,  
    boxsep=0pt,
    top=3pt,
    bottom=-3pt,
]
\begin{case}
Given the threat report:
"...Apache HTTP Server allows remote attackers to execute arbitrary code. The issue arises from improper handling of user-supplied input in HTTP request..." is related to CVE-2021-42013. LLMs present different outputs:
\begin{mitemize}
    \item GPT-o1 and DeepSeek-R1 incorrectly output CVE-2020-0609 and CVE-2022-2296, both of which are unrelated to Remote Code Execution or Apache.
    \item \system-70B gives CVE-2023-25690, a highly relevant HTTP Request Smuggling attack on the Apache Server.
\end{mitemize}
\end{case}
\end{tcolorbox}
From the above case study, although \system-70B does not match the exact CVE identifier, it effectively aligns with the evidence provided in the threat report and demonstrate a strong threat correlation capability. This effectiveness is further supported by the results shown in Figure \ref{fig:expt-3}-(c)(d), where \system presents stronger correlation relevance w.r.t. the correct products and threat categories.


{\bf RF\(_8\): \system demonstrates robust expansive correlation.}
Table \ref{tab:expt-4} presents the hit ratio for the top 10 expanded threats (CVEs) as reported in MITRE-CVE \cite{mitre-cve}. We evaluate three correlation strategies: (i) Direct expansion, (ii) CVEs as intermediates, and (iii) CWEs as intermediates. We have corresponding observations:
\begin{table}[!t]
  \small
  \def\arraystretch{0.95}
  \setlength{\tabcolsep}{3.5pt}
  \caption{Evaluation results (HIT@10) on expansive correlation task using historical threats (we do not evaluate on 0-day threats as they have incomplete reports of concurrent CVEs). The order of LLM names (left-to-right) follows the same presentation sequence as in Tables \ref{tab:expt-1} (top-down).}
  \begin{center}
  \scalebox{0.9}{
  \begin{tabular}{c|ccccccccccc}
     & {\bf SL} & {\bf LY} &{\bf 4O} & {\bf O1} & {\bf GM} & {\bf LA} & {\bf DS} & {\bf CD} & {\bf C1} & {\bf C8} & {\bf C70} \\
    \hline
    \hline
    Direct & 0.6 & 0.0 & 3.5 & 8.1 & 2.2 & 0.0 & 0.0 & 0.4 & 52.7 & \underline{74.3} & {\bf 77.5} \\
    w/ CVE & 7.3 & 4.5 & 11.6 & 25.4 & 6.7 & 0.0 & 4.3 & 12.5 & 60.4 & \underline{79.0} &  {\bf 81.2} \\
    w/ CWE & 0.0 & 0.0 & 0.0 & 0.8 & 0.0 & 0.0 & 0.0 & 0.0 & 36.2 & \underline{55.4} & {\bf 60.7}  \\
    \hline
  \end{tabular}
  }
  \end{center}
\label{tab:expt-4}
\end{table}
\begin{mitemize}
    \item[(i)] While baseline models largely struggle to expansively correlate CVEs, \system demonstrates significant effectiveness (e.g., 77.5\% HIT@10 with \system-70B). 
    \item[(ii)] Incorporating CVEs as intermediate evidence improves expansive correlation. However, baseline models tend to propagate erroneous CVEs from the prior detection step, which misguides subsequent threat correlations.
    \item [(iii)] Incorporating CWEs degrades expansive correlation due to their inherent ambiguity. Since a CWE can broadly relate to many CVEs, it introduces ambiguity, impeding accurate correlation. 
\end{mitemize}


\subsection{Threat Prioritization}

Real-world cyber threats often co-occur across cyber events \cite{rahman2022investigating,peng2020threat}, necessitating that CTI teams to distinguish threat severity and prioritize subsequent patches and mitigations. Our evaluation focuses on two prioritization measurements: severity level and analysis of exploitation dynamics.

{\bf Severity Level.} Threat severity can be assessed from threat reports even when CVE identifiers are not disclosed or when zero-day threats have not yet been assigned CVEs. We evaluate severity using 8 metrics defined in CVSS 3.x \cite{NVD_CVSSv3_Calculator}, which collectively capture a threat’s exploitability and impact. Each severity metric is treated as a classification task. For example, the {\it Attack Vector} metric has four predefined categories: {\it Network}, {\it Adjacent Network}, {\it Local}, and {\it Physical}. If a threat report lacks sufficient evidence for CVSS metrics, we introduce an additional label, "N/A", for each metric.

Furthermore, we classify the base severity level from four categories: Low, Medium, High, and Critical. This classification aggregates the ratings from eight CVSS metrics, providing an overall severity assessment aligned with CVSS standards.

{\bf RF\(_{9}\): \system maintains stable effectiveness.}
Table \ref{tab:expt-5} shows the accuracy of all models. While baselines vary widely across CVSS metrics (e.g., \textsc{GPT-o1} scores 92.91\% on Attack Complexity but only 20.43\% on Integrity Impact), \system remains consistently strong. Notably, \system-8B achieves $>$85\% accuracy across 8 metrics, demonstrating its reliability across different severity factors.

This stability comes from \system’s adaptability across different evidence types. Each CVSS metric relies on specific evidence, for example, Privileges Required (PR) depends on access control policies, while Attack Complexity (AC) is influenced by system constraints and attacker's skill. \system reduces inconsistencies in severity analyses by effectively integrating diverse evidence sources.



{\bf Exploitation Dynamics.} Besides severity analysis using CVSS, we also evaluate \system's effectiveness in analyzing dynamic exploitation behaviors. Specifically, we use the EPSS score \cite{EPSS2025} and conduct two types of analysis: (1) pinpointing historical exploitation at an earlier timestamp and (2) forecasting future exploitation at a later timestamp. Note that we exclude zero-day threats from the exploitation analysis due to the incompleteness of EPSS data.

\begin{figure}[h]
    \centering
    \includegraphics[width =80mm]{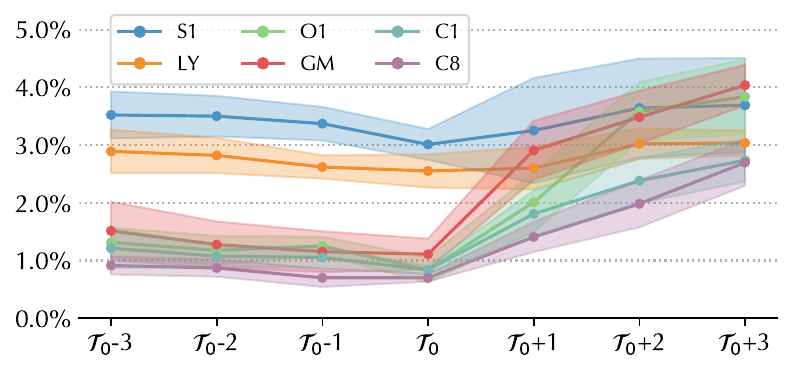}
    \caption{With $\mathcal{T}_0$ as the ``current'' timestamp (varying among threats), we compute RMSE for historical EPSS scores over the past three months (interpolation) and future EPSS scores for the next three months (prediction) w.r.t. $\mathcal{T}_0$.}
    \label{fig:exploit-trend}
\end{figure}

For test threats with more than six months of EPSS records, we set a reference timestamp $\mathcal{T}_0$ for each cyber threat (e.g., 01/01/2024) and collect EPSS reports from $\mathcal{T}_0$-3 months (e.g., 10/01/2023) to $\mathcal{T}_0$+3 months (e.g., 04/01/2024). However, EPSS reports between $\mathcal{T}_0$-3 to 
$\mathcal{T}_0$ are included in the inference prompt as prior knowledge, similar to the prompt structure illustrated in Figure \ref{fig:prompt}.

We compare \system-1B/8B against four baseline models and compute RMSE between the interpolated (historical) or predicted (future) EPSS scores and the actual scores (scaled from 0\% to 100\%). To ensure accurate evaluation, we match each EPSS score to the closest available recorded score. For example, if an EPSS score is not reported exactly on 10/01/2023, we use the closest available date (e.g., 09/25/2023) from EPSS reports \cite{mitre-cve} as the ground truth.

{\bf RF\(_{10}\): LLMs vary in interpolation and prediction.} As shown in Figure \ref{fig:exploit-trend}, LLMs perform slightly better in interpolating historical EPSS scores (lower RMSE) than in predicting future scores. However, we also observe that local LLMs, such as \textsc{SevenLLM}, perform obviously worse in both interpolation and prediction due to their static inherent knowledge, lacking adaptability to the emerging exploitation. On the contrary, \system performs similar to API-based LLMs (e.g., \textsc{Gemini}), which demonstrates the necessity of equipping LLMs with domain expertise. This enables LLMs to adapt to the dynamic nature of CTI tasks.

{\bf RF\(_{11}\): Contextual awareness is required.} We also observe that some cyber threats undergo dramatic shifts in EPSS scores due to emerging exploitation activities. This implies the importance of not only fitting to EPSS trends but also incorporating real-world events to enhance CTI capability. For instance:

\begin{tcolorbox}[
    colback=\boxcolor,  
    colframe=black, 
    width=0.48\textwidth,  
    boxrule=1pt, 
    arc=5pt,
    left=5pt,    
    right=5pt,  
    boxsep=0pt,
    top=3pt,
    bottom=2pt,
]
\begin{case}
CVE-2024-27956 presents a sharp increase in EPSS scores on 02/15/2025, jumping from 0.31\% to 55.37\% \cite{CVE2024-27956}. No LLM was able to accurately predict this trend without contextual knowledge of real-world events.
\end{case}
\end{tcolorbox}

\begin{figure*}[!h]
    \centering
    \includegraphics[width=\textwidth]{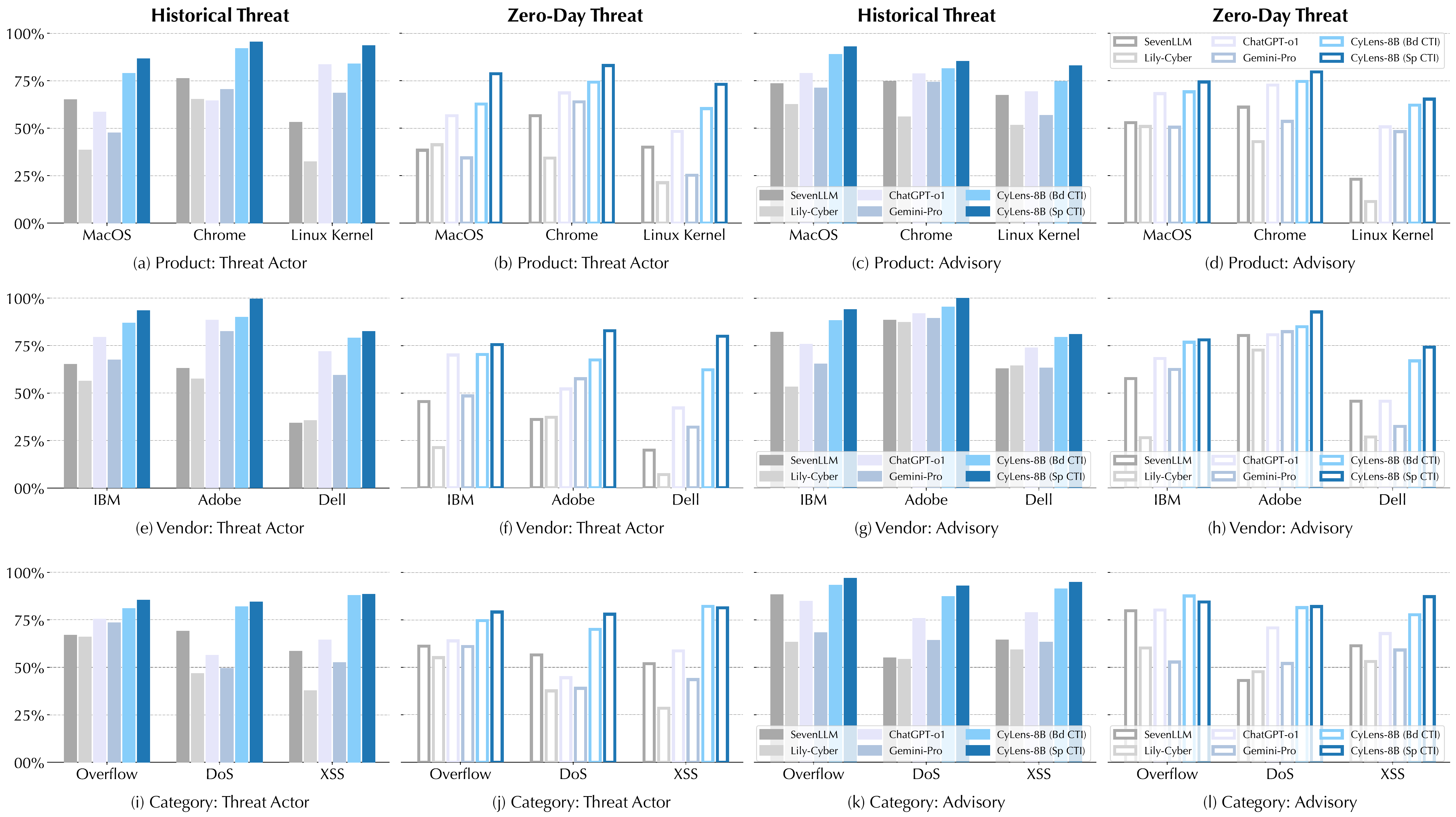}
    \caption{Evaluation results by different groups of threats: (a)–(d) correspond to threats targeting the same product, (e)–(h) to those affecting the same vendor, and (i)–(l) to threats within the same category. We further divide these threats into historical and zero-day groups, with dataset statistics provided in Table \ref{tab:eval-data-stat}. The evaluation metrics used are accuracy for both threat actor attribution and third-party advisory suggestions. Abbrevation: "Bd CTI" -- \system fine-tuned by broad CTI tasks (same setting as \cref{sec:expt}); "Sp CTI" -- \system fine-tuned by specialized CTI tasks. Additional evaluation results are shown in Figure \ref{fig:organize-2}.}
    \label{fig:organize}
\end{figure*}

\subsection{Remediation}

\begin{table}[!t]
  \small
  \def\arraystretch{0.8}
  \setlength{\tabcolsep}{4.8pt}
  \caption{Evaluation results on {\bf \ding{187} Remediation}. We evaluate whether LLMs can correctly generate: (i) patching or mitigation tools -- measured by accuracy, (ii) code patches -- measured by syntactic similarity, (iii) patching or mitigation methods -- measured by BERT Score, and (iv) reference advisories -- measured by accuracy.}
  \begin{center}
  \scalebox{0.90}{
  \begin{tabular}{ccccccccc}
    \toprule
     \multirow{2.5}{*}{\bf Model} & \multicolumn{2}{c}{\bf Tool Use} & \multicolumn{2}{c}{\bf Code Patch} & \multicolumn{2}{c}{\bf Methodology} & \multicolumn{2}{c}{\bf Advisory} \\
    \cmidrule(lr){2-3} \cmidrule(lr){4-5}\cmidrule(lr){6-7}\cmidrule(lr){8-9} 
    & Hist. & \hlcell 0-Day & Hist. & \hlcell 0-Day & Hist. & \hlcell 0-Day & Hist. & \hlcell 0-Day\\
    \midrule
    SL & 87.81 & \hlcell 79.65 & 81.83 & \hlcell 76.62 & 82.95 & \hlcell 82.48 & 56.74 & \hlcell 43.25 \\
    LY & 79.63 & \hlcell 57.17 & 75.96 & \hlcell 73.60 & 80.75 & \hlcell 78.53 & 48.71 & \hlcell 46.63 \\
    \midrule
    4O & 86.62 & \hlcell 74.51 & 82.70 & \hlcell 81.05 & 84.42 & \hlcell 82.51 & 76.83 & \hlcell 71.75 \\
    O1 & 90.08 & \hlcell 86.65 & 84.47 & \hlcell 82.86 & 85.53 & \hlcell 85.25 & 81.26 & \hlcell 77.79 \\
    GM & 76.83 & \hlcell 53.22 & 83.14 & \hlcell 82.47 & 84.79 & \hlcell 80.32 & 66.72 & \hlcell 43.25 \\
    LA & 75.71 & \hlcell 69.20 & 81.23 & \hlcell 78.46 & 83.26 & \hlcell 82.59 & 49.53 & \hlcell 38.17 \\
    DS & 73.70 & \hlcell 70.21 & 79.27 & \hlcell 77.51 & 83.64 & \hlcell 82.55 & 65.43 & \hlcell 47.92 \\
    CD & 77.54 & \hlcell 73.48 & 76.19 & \hlcell 73.64 & 82.27 & \hlcell 81.92 & 54.85 & \hlcell 47.35 \\
    \midrule
    C1 & 92.03 & \hlcell 88.84 & 85.42 & \hlcell 83.71 & 87.83 & \hlcell 84.52 & 88.74 & \hlcell 82.25 \\
    C8 & \underline{93.83} & \underline{\hlcell 92.74} & \underline{87.48} & \hlcell {\bf 87.36} & \underline{89.16} & \hlcell {\bf 88.37} & \underline{92.32} & \hlcell \underline{88.54} \\
    C70 & {\bf 97.62} & \hlcell {\bf 97.23} & {\bf 90.66} & \hlcell \underline{86.39} & {\bf 91.20} & \hlcell \underline{88.04} & {\bf 95.34} & \hlcell {\bf 92.26} \\
    \bottomrule
  \end{tabular}
  }
  \end{center}
\label{tab:expt-6}
\end{table}

Finally, we evaluate \system's capability in generating accurate remediation strategies, focusing on four aspects: suggesting the correct tool for patching or mitigation (e.g., OpenSSL for encryption), generating available code patches (e.g., GitHub commits), proposing appropriate mitigation methodologies (e.g., hardening an OS), and identifying the correct advisory to consult (e.g., NVIDIA Security Advisory).

{\bf RF\(_{12}\):  Broad expertise is crucial.} Consistent with RF\(_{1}\) and RF\(_{9}\), we demonstrate \system's effectiveness in addressing a wide range of remediation objectives, as shown in Table \ref{tab:expt-6}. These include both identification-based tasks (e.g., suggesting appropriate tools and advisories) and generation-based tasks (e.g., producing code patches or mitigation strategies). In contrast, baseline models often exhibit imbalanced performance across remediation tasks. For instance, while \textsc{Gemini-Pro} performs well in generating code patches, it struggles to correctly identify relevant advisories.

We attribute this disparity to the fact that effective remediation often depends on familiarity with established cybersecurity practices \cite{jacobs2020improving}. As such, embedding large-scale historical remediation knowledge is essential when developing LLMs for CTI applications. Notably, even for zero-day threats, a well-trained model like \system can leverage its inherent domain knowledge to derive meaningful insights, thereby enhancing response effectiveness in scenarios where no explicit remediation guidance exists.

\section{Evaluation Results: Specialized CTI}
\label{sec:expt-adapt}

Despite \textsc{CyLens}’s scalability in capturing a broad threat landscape, real-world CTI often involves targeted interests \cite{mvula2024survey,optiv_threat_services}. In this section, we demonstrate \system's adaptability in addressing three specialized groups of threats:
\begin{mitemize}
    \item {\bf Product} -- Focusing on cyber threats affecting specific products, a common scenario among security teams in large enterprises managing product-specific CTI \cite{cisa_cyber_essentials}. 
    \item {\bf Vendor}  -- Targeting threats that impact an entire vendor (i.e., company), which is a common practice among industry groups, particularly those have focused products or services.\cite{optiv_threat_services}.
    \item {\bf Category} -- Concentrating on specific threat categories, which is particularly relevant to research institutions conducting specialized cybersecurity studies \cite{kim2022threat}.
\end{mitemize}

{\bf Specialized Instruction Tuning.} Following the same re-training process described in \cref{ssec:pt}, \system is {\bf independently} fine-tuned on threats associated with specific products (MacOS, Chrome, or Linux Kernel), vendors (IBM, Adobe, or Dell), and categories (Overflow, DoS, or XSS). These selections are based on their exploitation popularity as recorded in the MITRE-CVE database \cite{cve}. The fine-tuning datasets are derived from a subset of the original collections (Table \ref{tab:ft-stat}) that include only historical threats, with no overlap with the evaluation dataset (Table \ref{tab:eval-data-stat}). 

{\bf Evaluation Targets.} As detailed in \cref{sec:expt-setting}, we evaluate \system using real-world threat reports collected from multiple platforms \cite{oracle_security, redhat_bugzilla, adobe_security_bulletin, apache_mailing_list, debian_security, opensuse_security, fedora_security}. From these sources, we extract threats targeting the selected products, vendors, and categories. The corresponding dataset statistics are reported in Table \ref{tab:eval-data-stat}.

We compare the specialized \system-8B model with two SOTA, \textsc{SevenLLM} and \textsc{Lily-Cyber}, and two industry-leading LLMs, \textsc{GPT-o1} and \textsc{Gemini}. Additionally, we include a non-specialized \system-8B fine-tuned on a broad range of CTI tasks under the same settings as in \cref{sec:expt}. The evaluation results across multiple CTI tasks are summarized in Figure \ref{fig:organize} and \ref{fig:organize-2}. We have the following findings:

{\bf RF\(_{13}\): \system is adaptable to diverse specialized needs.} From Figure \ref{fig:organize}, we observe that \system maintains high effectiveness even with independent instruction-tuning across different groups of cyber threats. For instance, \system-8B achieves nearly 100\% accuracy in providing remediation advisories for Adobe. This result further demonstrates \system's adaptability despite the diverse nature of threats across products, vendors, and categories—factors that significantly impact baseline LLMs, causing their performance to fluctuate. 


{\bf RF\(_{14}\):  Data scarcity matters.} Cyber threats targeting specific products or even entire vendors often suffer from data scarcity, particularly a lack of historical threats that are important for experience-based remediation. This data scarcity can hinder the development of countermeasures, especially for zero-day threats. For example:

\begin{tcolorbox}[
    colback=\boxcolor,  
    colframe=black, 
    width=0.48\textwidth,  
    boxrule=1pt, 
    arc=5pt,
    left=5pt,    
    right=5pt,  
    boxsep=0pt,
    top=3pt,
    bottom=2pt,
]
\begin{case}
As of 2024, only 9 cyber threats related to Open Redirection vulnerabilities had been reported for Dell. However, within the first three months of 2025, 2 new threats emerged—targeting Dell Unity (CVE-2025-24381) and Networker (CVE-2025-21104)—highlighting the challenge of generating timely and effective mitigation strategies with limited historical data.
\end{case}
\end{tcolorbox}

{\bf RF\(_{15}\): A broad view of the threat landscape is necessary.}
When developing CTI solutions for specific products or vendors, relying solely on threat data limited to those entities is often insufficient (as highlighted in RF\(_{14}\)). It is essential to consider the broader threat landscape to uncover relevant analytical patterns or remediation strategies. This is demonstrated by the effectiveness of \system—both \system-1B and \system-8B achieve significantly higher accuracy in advisory suggestion for Dell-related threats (>70\%) compared to all baseline methods (<50\%). A representative case study is:

\begin{tcolorbox}[
    colback=\boxcolor,  
    colframe=black, 
    width=0.48\textwidth,  
    boxrule=1pt, 
    arc=5pt,
    left=5pt,    
    right=5pt,  
    boxsep=0pt,
    top=3pt,
    bottom=2pt,
]
\begin{case}
In studying the remediation of CVE-2025-21104, \system-8B successfully identifies that reconfiguring the web server to enforce strict parsing rules and ignore unexpected Host header values can mitigate the vulnerability—an approach typically incorporated in newer system versions. This insight aligns with Dell's official advisory DSA-2025-124, which includes a security update for the Dell NetWorker Management Console addressing the issue.
\end{case}
\end{tcolorbox}

\section{Discussion}

\subsection{Ablation Study}
\label{ssec:ablation}

We conduct ablation studies to assess the impact of different components within \system. Table \ref{tab:ablation} presents the results, where we independently exclude re-training (RT), fine-tuning (FT), and four functional modules: TOM, NER, REL, and RAG. Notably, REA and SUM cannot be excluded, as they are essential inference steps.

\begin{table}[h]
  \small
  \def\arraystretch{0.85}
  \setlength{\tabcolsep}{4.5pt}
  \caption{Ablation study on \system-8B by selecting four representative CTI tasks: (1) attributing threat actors (measured by accuracy), (2) contextualizing attack infrastructure (measured by F1 score), (3) classifying CVSS base severity levels (measured by accuracy), and (4) suggesting advisories for remediation (measured by accuracy).}
  \begin{center}
  \scalebox{0.9}{
  \begin{tabular}{ccccccccc}
    \toprule
     \multirow{2.5}{*}{\bf Ablation} & \multicolumn{2}{c}{\bf Threat Actor} & \multicolumn{2}{c}{\bf Attack Infra} & \multicolumn{2}{c}{\bf CVSS Base} & \multicolumn{2}{c}{\bf Advisory}\\
    \cmidrule(lr){2-3} \cmidrule(lr){4-5}\cmidrule(lr){6-7}\cmidrule(lr){8-9}
    & Hist. & \hlcell 0-Day & Hist. & \hlcell 0-Day & Hist. & \hlcell 0-Day & Hist. & \hlcell 0-Day\\
    \midrule
    w/o RT & 66.22 & \hlcell 62.31 & 70.05 & \hlcell 61.44  & 53.62 & \hlcell 45.90 & 57.25 & \hlcell 54.36 \\
    w/o FT & 82.42 & \hlcell 70.15 & 86.47 & \hlcell 78.31  & 86.48 & \hlcell 61.75 & 82.56 & \hlcell 77.23 \\
    \midrule
    w/o TOM & 84.23 & \hlcell 81.17 & 88.46 & \hlcell 82.24  & 95.12 & \hlcell 88.82 & 87.23 & \hlcell 85.44 \\
    w/o NER & 71.12 & \hlcell 70.19 & 82.83 & \hlcell 72.63  & 94.52 & \hlcell 91.42 & 82.53 & \hlcell 80.65 \\
    w/o REL & 87.21 & \hlcell 83.09 & 88.18 & \hlcell 86.21  & 95.67 & \hlcell 91.25 & 83.12 & \hlcell 82.72 \\
    w/o RAG & 84.17 & \hlcell 80.26 & 86.53 & \hlcell 82.71  & 95.23 & \hlcell 89.64 & 85.14 & \hlcell 83.42 \\
    \bottomrule
  \end{tabular}
  }
  \end{center}
\label{tab:ablation}
\end{table}

We observe that re-training has the most significant impact, as it embeds massive CTI knowledgeinto the model. While fine-tuning is more lightweight, its influence remains non-negligible in refining \system’s CTI capabilities.
Additionally, the impact of different modules varies across tasks. For instance, classifying CVSS severity levels does not heavily depend on NER and REL, so removing these modules has minimal effect. However, tasks such as threat actor attribution or advisory suggestion rely on accurately identifying key objects (e.g., malware or third-party vendors), making NER and REL critical for performance.
These findings reinforce the necessity of incorporating all proposed techniques to achieve an effective and well-rounded system design for CTI applications.

\subsection{Adapting to Computational Constraints}

Similar to Meta's Llama series, which offers models at different scales \cite{llama3.1-8b,llama3.2-11b,touvron2023llama}, \system is available in three sizes: 1B, 8B, and 70B, allowing it to accommodate varying computational constraints. While \system-1B is less effective compared to the 8B and 70B, its lightweight design is crucial for resource-limited deployments.

In scenarios where computational resources are highly constrained, having a functional LLM is often more practical than striving for state-of-the-art performance. Notably, despite its smaller size, \system-1B still outperforms industry-leading LLMs in most CTI tasks, reinforcing its value as an efficient yet capable model for cybersecurity applications.

\section{Conclusion}
\label{sec:conclusion}

We introduce \system, a LLM-powered agentic CTI system. \system supports the entire threat management lifecycle, encompassing tasks from threat attribution to remediation. It accumulates CTI knowledge from massive threat reports and utilizes a functional module for real-time analysis, enabling it to adeptly handle both historical and zero-day threats. Intensive evaluations have demonstrated the scalability and adaptability of \system across massive CTI tasks and tailored needs of threat management. By consistently outperforming industry-leading LLMs and SOTA agents, \system offers scalable and adaptive solution and serves as a blueprint for integrating agentic LLMs into CTI practice.

\newpage
\bibliographystyle{ACM-Reference-Format}
\bibliography{reference}
\label{reference}

\newpage
\appendix
\newpage
\section{Pre-Training: Complementary Information}
\label{app:pt}

\begin{table*}[!t]
  \small
  \setlength{\tabcolsep}{5pt}
  \caption{The data sources we utilize and the corresponding evidence metadata collected from each database.}
  \begin{center}
  \begin{tabular}{ccccccccc}
    \toprule
    \multirow{2}{*}{\bf CTI Task} & \multirow{2}{*}{\bf Analysis Target} & \multicolumn{7}{c}{\bf Source Database} \\
    \cmidrule(lr){3-9}
    & & CVE & NVD & CWE & ATT\&CK & CAPEC & Exploit DB & Third-Party Ref. \\ 
    \midrule
    \multirow{3}{*}{\ding{182}} & Threat Actor & \cmark & \cmark & \cmark  & \cmark  &  & \cmark & \cmark \\
    & TTPs & \cmark & \cmark & & \cmark &  & \cmark  & \cmark \\
    & Campaign & \cmark & \cmark & \cmark  & \cmark & \cmark & \cmark & \cmark  \\
    \cmidrule(lr){2-9}
    \multirow{3}{*}{\ding{183}} & Affected System & \cmark & \cmark & & \cmark  & & \cmark & \cmark  \\
    & Attack Infra & \cmark & \cmark & &  & \cmark  & \cmark & \cmark \\
    & Attack Impact & \cmark & \cmark & &  & & \cmark & \cmark \\
    \cmidrule(lr){2-9}
    \multirow{2}{*}{\ding{184}} & CVE Identifier & \cmark & \cmark  &  &  & & \cmark  & \cmark  \\
    & CWE Identifier & \cmark & \cmark & \cmark & & & & \cmark  \\
    \cmidrule(lr){2-9}
    \multirow{2}{*}{\ding{185}} & CVSS Metrics  & \cmark  & \cmark  & & & & & \cmark \\
    & EPSS Records & \cmark & \cmark  & &  &  & & \cmark \\
    \cmidrule(lr){2-9}
    \multirow{4}{*}{\ding{186}} & Tool Use & \cmark & \cmark  & \cmark & \cmark  &  & \cmark & \cmark \\
    & Code Patch & \cmark & \cmark  & &  &  & \cmark & \cmark \\
    & Methodology & \cmark & \cmark & \cmark  & \cmark  &  & \cmark & \cmark \\
    & Advisory & \cmark & \cmark  & \cmark  &  \cmark &  & \cmark & \cmark \\
    \bottomrule
  \end{tabular}
  \end{center}
\label{tab:source}
\end{table*}

\subsection{Data Source and Metadata Collection}
\label{app:source}

Complementary to $\cref{ssec:pt-data}$, Table \ref{tab:source} provides a detailed overview of the cyber threat databases we utilize, along with the corresponding metadata collected for various CTI tasks (\ding{182}-\ding{186}) and analytical targets.

{\bf MITRE-CVE and NVD.} Our collected data is primarily centered around CVEs, leveraging the MITRE-CVE \cite{mitre-cve} and NVD \cite{nvd} databases, which serve as well-structured repositories of publicly disclosed vulnerabilities. These databases provide foundational information, including CVE identifiers, descriptions, severity scores, affected products, and references to external sources.

{\bf Exploit-DB.} To further enrich our dataset and enhance contextual understanding, we incorporate additional cyber threat intelligence (CTI) sources, including Exploit-DB \cite{exploit-db}. Exploit-DB provides reports on security advisories and threat intelligence insights, complementing the structured data from MITRE-CVE and NVD. Notably, Exploit-DB is also CVE-centric; however, unlike MITRE-CVE and NVD, it lacks structured connections to additional evidence. Therefore, our data collection primarily focuses on extracting raw threat descriptions and attack vector classifications (e.g., "Local") to supplement our dataset with additional contextual details.

{\bf Other MITRE Databases.} Beyond these primary databases, we integrate metadata from additional sources to enrich our dataset with diverse perspectives on cyber threats. Specifically, we leverage CWE \cite{cwe} to systematically categorize software weaknesses, providing a structured foundation for understanding vulnerability root causes. CAPEC \cite{capec} offers a comprehensive taxonomy of attack patterns, enabling us to analyze how adversaries exploit known weaknesses in real-world scenarios. MITRE ATT\&CK \cite{mitre-attack} serves as a crucial resource for mapping adversary tactics, techniques, and procedures (TTPs), facilitating in-depth attribution and behavioral analysis.

{\bf Third-Party References.} Beyond structured frameworks, we incorporate third-party references to capture real-time threat intelligence and contextualized security insights. Among them, GitHub commits to serve as a valuable source for tracking vulnerability disclosures, exploit development, and security patch discussions. Apache security reports \cite{ApacheFoundation} provide detailed documentation on vulnerabilities and remediation efforts specific to widely used web and enterprise software. Meanwhile, IBM X-Force threat intelligence \cite{IBM_XForce} contributes curated threat reports, real-world attack observations, and insights into evolving threat landscapes.

The collected metadata, detailed in Table \ref{tab:source}, enhances the comprehensiveness of our dataset by offering multifaceted insights into vulnerability exploitation patterns, adversarial attack techniques, and effective mitigation strategies. By integrating these diverse sources, we enable a richer and more contextualized analysis of cyber threats, supporting improved detection, attribution, and defense mechanisms.

\subsection{Threat Corpus Generation Details}
\label{app:pt-gen-detail}

Complementary to \cref{ssec:pt-data}, we detail the prompts used for generating and refining the threat corpus, specifically in Prompt \ref{prompt:generation} and \ref{prompt:revision}. During output sampling, we adjust the temperature and top-p parameters within a range of 0.2 to 1.0 to balance creativity and determinism. An example of a generated threat report is provided in Example \ref{example:threat-report}. Additionally, we present the statistical overview of the corpus used for pre-training in Table \ref{tab:stat}.

\begin{prompt}
Below is the prompt template used for threat corpus generation.\\
\fbox{\parbox{\dimexpr\linewidth-2\fboxsep-2\fboxrule\relax}{
\label{prompt:generation}
{\bf Task Description:}
You are a cybersecurity analyst responsible for generating a high-quality CVE-centric threat report. The report should follow the style, layout, and analytical depth of the given example report. $\cdots$

\vspace{5pt}
{\bf Example Report:}
\{EXAMPLE REPORT\}  

\vspace{5pt}
{\bf Threat Metadata:}  
\begin{mitemize}
    \item {\bf Vulnerability Name:}  \{VULNERABILITY NAME\}
    \item {\bf CVE ID:} \{CVE ID\}
    \item {\bf Affected Systems:} \{AFFECTED SYSTEMS\}
    \item {\bf Attack Vector:} \{ATTACK VECTOR\}
    \item {\bf Tactics, Techniques, and Procedures (TTPs):} \{TTPS\}
    \item {\bf Known Exploitations:} \{KNOWN EXPLOITATIONS\}
    \item $\cdots$
\end{mitemize}

\vspace{5pt}
{\bf Report Generation Instructions:}
\begin{mitemize}
    \item Follow the style and structure of the example report.
    \item Incorporate all CVE metadata meaningfully.
    \item Analyze the threat context, discussing possible exploitation paths, attacker motivations, and mitigation methods.
    \item Ensure coherence while maintaining variability in phrasing.
    \item Generate the report in professional cybersecurity language while maintaining clarity and depth.
\end{mitemize}

\vspace{5pt}
{\bf Output:}
}}
\end{prompt}

\begin{prompt}
Below is the prompt used for threat corpus revision.\\
\fbox{\parbox{\dimexpr\linewidth-2\fboxsep-2\fboxrule\relax}{
\label{prompt:revision}
{\bf Objective:} You are an expert cybersecurity analyst and technical writer. Your task is to revise the provided cyber threat report by improving its layout and structure while maintaining all critical details. The goal is to $\cdots$

\vspace{3pt}
{\bf Instructions:}
\begin{mitemize}
    \item Reorganize Content – Adjust the structure to make the report more logical and easier to follow. Consider using sections such as $\cdots$
    \item Improve Readability – Use concise paragraphs, bullet points, and tables where appropriate. Ensure technical jargon is well-explained.
    \item Maintain Accuracy – Preserve all details, technical data, and intelligence without altering their meaning.
    \item $\cdots$
\end{mitemize}

\vspace{3pt}
{\bf Report to Revise:}
[Insert Cyber Threat Report Here]

\vspace{3pt}
{\bf Output:}
}}
\end{prompt}

\begin{example}
Below is an example of threat report.\\
\fbox{\parbox{\dimexpr\linewidth-2\fboxsep-2\fboxrule\relax}{
\label{example:threat-report}
{\bf Title:} Security Advisory: Cleartext Storage Vulnerability in ChatGPT macOS Application 

\vspace{5pt}
{\bf Executive Summary:} A security vulnerability has been identified in versions of the ChatGPT application for macOS released before July 5, 2024. The application was found to store user conversations in cleartext, making them accessible to other applications on the system. Additionally, the app operated outside of the macOS sandbox, $\cdots$

\vspace{5pt}
{\bf Vulnerability Details:}
\begin{mitemize}
    \item {\bf CVE Identifier:} CVE-2024-40594
    \item {\bf Description:} The ChatGPT macOS application prior to version 2024-07-05 stored user conversations in an unencrypted format in a location accessible by other applications. Furthermore, the application opted out of the macOS sandbox, which is designed to limit the app's access to system resources and user data $\cdots$
    \item {\bf Affected Products:} $\cdots$
\end{mitemize}

\vspace{5pt}
{\bf Impact:} The cleartext storage of conversations poses a privacy risk, as sensitive information could be accessed by other applications or users with sufficient privileges on the same system. The lack of sandboxing further exacerbates the issue by allowing the application broader access to system resources than necessary, potentially exposing additional data $\cdots$

\vspace{5pt}
{\bf Mitigation:} Users are advised to update the $\cdots$

}}
\end{example}


\section{Instruction Tuning: Complementary Detail}
\label{app:ft}

Complementary to \cref{ssec:ft}, we present the statistics of the generated instruction-tuning dataset in Table \ref{tab:ft-stat}. Notably, when generating each data instance using a cascading reasoning layout, not all analytical targets are included due to variations in their availability.

\begin{table}[h]
  \small
  \def\arraystretch{1.0}
  \setlength{\tabcolsep}{5pt}
  \caption{Statistics of the instruction-tuning dataset with a cascading reasoning layout, where each analytical target appears in varying volumes of data. Additionally, different scales of data are used for different \system models to align with their respective capacities and optimization needs.}
  \begin{center}
  \scalebox{0.9}{
  \begin{tabular}{cccccc}
    \toprule
    \multirow{2.5}{*}{\bf CTI Task} &  \multirow{2.5}{*}{\bf Analytical Target} & \multirow{2.5}{*}{\bf \# Sample}  & \multicolumn{3}{c}{\bf \# Fine-Tuning Samples}\\
    \cmidrule(lr){4-6}
    & & & 1B & 8B & 70B \\
    \midrule
    \multirow{3}{*}{\ding{182}} & Threat Actor & 37.6K & 1.2K & 11.5K & 37.6K \\
    & TTPs & 35.5K & 1.2K & 11.3K & 35.5 K \\
    & Campaign & 21.8K & 0.8K & 7.9K & 21.8K \\
    \cmidrule(lr){2-6}
    \multirow{3}{*}{\ding{183}} & Affected System & 37.6K & 1.2K & 11.5K & 37.6K \\
    & Attack Infra & 28.3K & 0.9K & 10.2K & 28.3K \\
    & Attack Impact & 17.6K & 0.6K & 5.4K & 17.6K \\
    \cmidrule(lr){2-6}
    \multirow{2}{*}{\ding{184}} & CVE Identifier & 37.6K & 1.2K & 11.5K & 37.6K \\
    & CWE Identifier & 37.6K & 1.2K & 11.5K & 37.6K \\
    \cmidrule(lr){2-6}
    \multirow{2}{*}{\ding{185}} & CVSS Metrics & 37.6K & 1.2K & 11.5K & 37.6K \\
    & EPSS Records & 33.1K & 1.3K & 12.6K & 33.1K  \\
    \cmidrule(lr){2-6}
    \multirow{4}{*}{\ding{186}} & Tool Use & 15.3K  & 0.7K & 4.8K & 15.3K  \\
    & Code Patch & 8.2K & 1.1K & 3.1K & 8.2K \\
    & Methodology & 32.4K & 1.0K & 14.8K & 32.4K \\
    & Advisory & 27.5K & 0.8K & 9.6K  & 27.5K \\
    \bottomrule
  \end{tabular}}
  \end{center}
\label{tab:ft-stat}
\end{table}

Furthermore, \system-1B, \system-8B, and \system-70B utilize different volumes of fine-tuning data, reflecting their respective model capacities and optimization requirements. For fine-tuning \system-1B, we select a smaller volume of samples that contain all analytical targets. In contrast, for \system-8B and \system-70B, we use a progressively larger dataset, with \system-8B utilizing a partial subset and \system-70B leveraging the full dataset, despite the potential absence of certain analytical targets in some instances.

\section{Inference Module: Complementary Detail}
\label{app:nlp}

This section detail the implementation of six NLP modules as described in \cref{ssec:inference}.

\subsection{Topic Modeling}

To perform topic modeling, we leverage \texttt{BERTopic}, a SOTA topic modeling framework that integrates transformer-based embeddings with clustering and class-based TF-IDF. In our implementation, we design a two-stage prompt-based workflow to integrate BERTopic with \system for dynamic topic modeling. In the first stage, \system is prompted to assess whether topic modeling is needed based on the input text’s characteristics. \system receives instructions: 

\begin{prompt} Necessity of Topic Modeling. \\ 
\fbox{\parbox{\dimexpr\linewidth-2\fboxsep-2\fboxrule\relax}{
Given the input query, determine whether topic modeling is appropriate. 
    
Hint: topic modeling is suitable when the texts cover multiple themes or the user seeks to explore underlying topics.
}}
\end{prompt}

\system evaluates the semantic content and returns a binary decision along with reasoning. Once \system determines that topic modeling is appropriate, it proceeds to construct a structured call to \texttt{BERTopic} via a second-stage prompt. This prompt guides \system to assemble the necessary inputs for \texttt{BERTopic} and invoke it accordingly. The prompt is: 

\begin{prompt} Calling Topic Modeling. \\ 
\fbox{\parbox{\dimexpr\linewidth-2\fboxsep-2\fboxrule\relax}{
You have identified that topic modeling is required. Use \texttt{BERTopic} to extract topics from the following documents. For each topic, return a descriptive label, a list of representative keywords, and top exemplar documents.
}}
\end{prompt}

\system structures this as a function call or prompt completion depending on the system design. \texttt{BERTopic} is then executed with the provided documents, and the results are passed back to \system, which can further head to the next NLP module.

\subsection{Named Entity Recognition (NER)}
\label{app:ner-detail}

To perform NER, we adopt a prompt-based inference approach in \system in place of traditional sequence labeling methods. Rather than fine-tuning, \system is queried at runtime with carefully crafted prompts to identify and categorize named entities directly from the input text. The prompt is

\begin{prompt}  NER Prompt (An Example for Attribution Task). \\ 
\fbox{\parbox{\dimexpr\linewidth-2\fboxsep-2\fboxrule\relax}{
{\bf System Prompt:}

You are a cybersecurity threat intelligence assistant specialized in named entity recognition (NER). Your task is to extract and categorize entities relevant to threat attribution from unstructured text. Focus on identifying entities that help answer the question "Who is responsible for the attack?" and "How was the attack carried out?"

{\bf Instructions:}

Given a cybersecurity-related document or report excerpt, extract all relevant named entities and classify them into the following categories:

\begin{mitemize}
    \item Threat Actor: Names or aliases of individuals or groups suspected or known to be behind the activity.
    \item $\cdots$
\end{mitemize}

Return the results in the following structured JSON format.
}}
\end{prompt}

\system then returns a list of entities, each annotated with its type and surface form. 

\subsection{Relation Extraction}
\label{app:rel-detail}

To extract semantic relationships between entities in cybersecurity texts, we implement relation extraction through prompt-based inference in \system. This method allows us to bypass traditional supervised training by leveraging the \system’s few-shot and zero-shot reasoning capabilities. At inference time, we provide the \system with structured prompts that instruct it to identify and describe relationships between extracted entities, such as connections between threat actors and malware, or between attack campaigns and their target sectors. A typical prompt format is

\begin{prompt} Relation Extraction Prompt \\ 
\fbox{\parbox{\dimexpr\linewidth-2\fboxsep-2\fboxrule\relax}{
{\bf System Prompt:}

You are a cybersecurity threat intelligence reasoning assistant. Your task is to perform relation extraction to identify and structure meaningful semantic relationships between entities mentioned in cybersecurity texts, particularly in the context of threat attribution. The goal is to infer who did what to whom, how, when, and using what tools or infrastructure.

{\bf Instructions:}

Given a cybersecurity-related passage (such as a threat report or incident summary) and its extracted named entities, identify relevant relationships between entities and represent them as a list of (subject, relation, object) triples. Focus on attribution-relevant relations. Use only clearly supported relationships based on the input. If no relationships are found, return an empty list.

Return the results in this structured JSON format.

}}
\end{prompt}

\subsection{Retrieval-Augmented Generation (RAG)}

To enhance the quality of \system inference, we implement a RAG framework that integrates external web-based APIs during inference. This bridges generative reasoning with real-time access to structured and unstructured data sources, such as open-source intelligence feeds, public CTI repositories, and search APIs. 

The RAG workflow begins with the \system analyzing the user’s input to determine whether external retrieval is necessary. For example, if the input question involves current threat actor activity, recent campaign data, or infrastructure details (e.g., domains, malware hashes), the system triggers a retrieval phase. In this phase, \system is prompted to generate a structured search query that captures the essential information needs of the input. The prompt format follows a declarative schema, such as: {\it “Formulate a concise search query to retrieve current information about [topic], including actors, tools, targets, and known affiliations.”} The generated query is sent to a relevant external data source via API—for instance, querying public CTI portals (e.g., MITRE ATT\&CK, VirusTotal, Shodan, or threat feed aggregators) or web search APIs (e.g., Bing Search or Google Custom Search).

Upon receiving the retrieved content (typically one or more passages or structured documents), the system constructs a new input prompt to \system, embedding the original question and the retrieved evidence into the context window. This second-stage prompt is crafted using templates such as: {\it “Given the following query and retrieved documents, generate a comprehensive, evidence-based answer. Use only the provided content.”} This constraint guides \system to ground its reasoning in the externally retrieved information, thereby improving factual accuracy and traceability.

\subsection{Reasoning}

To support advanced analytical capabilities in \system, we implement modular reasoning and summarization components using prompt engineering. Rather than relying solely on fine-tuned models for specific downstream tasks, our architecture employs LLMs (i.e., \system itself) as general-purpose reasoning engines, guided by carefully constructed prompts. This allows \system to remain flexible, easily extensible, and capable of performing domain-specific tasks such as cyber threat correlation, threat actor profiling, and intelligence synthesis without retraining.

The reasoning module is designed to interpret and connect extracted entities, attributes, and contextual cues across threat intelligence documents. We use prompt engineering to simulate multi-step analytical thinking, instructing the LLM to infer relationships, fill in missing links, and assess causality. For instance, prompts are constructed as: {\it “Given the following entities and attack context, identify the most plausible attribution and justify your reasoning step-by-step.”} This encourages the model to break down the logic behind associations, such as linking a threat actor to a campaign based on observed TTPs, tool reuse, or geopolitical motivation. We also use chain-of-thought prompting and structured templates (e.g., tabular reasoning) to improve traceability and consistency of the model’s analytical conclusions.

\subsection{Summarization}

The summarization module in \system is responsible for generating high-level, human-readable briefings from dense technical threat reports or multi-source intelligence inputs. To achieve this, we implement domain-adapted summarization prompts that guide \system to preserve critical cybersecurity-relevant information such as named entities, TTPs, indicators of compromise (IOCs), and timeline markers. The prompt format typically follows: “Summarize the following document into a concise threat intelligence report, highlighting threat actors, attack vectors, affected sectors, and key dates.” Depending on the use case, we support multiple summarization styles, including executive summaries, technical briefs, and IOC-centric extracts. We also integrate few-shot examples when needed to adapt the tone or structure to a specific user role (e.g., SOC analyst vs. CISO).

\section{Evaluation: Complementary Information}
\label{app:eval}

\begin{figure*}[!h]
    \centering
    \includegraphics[width=\textwidth]{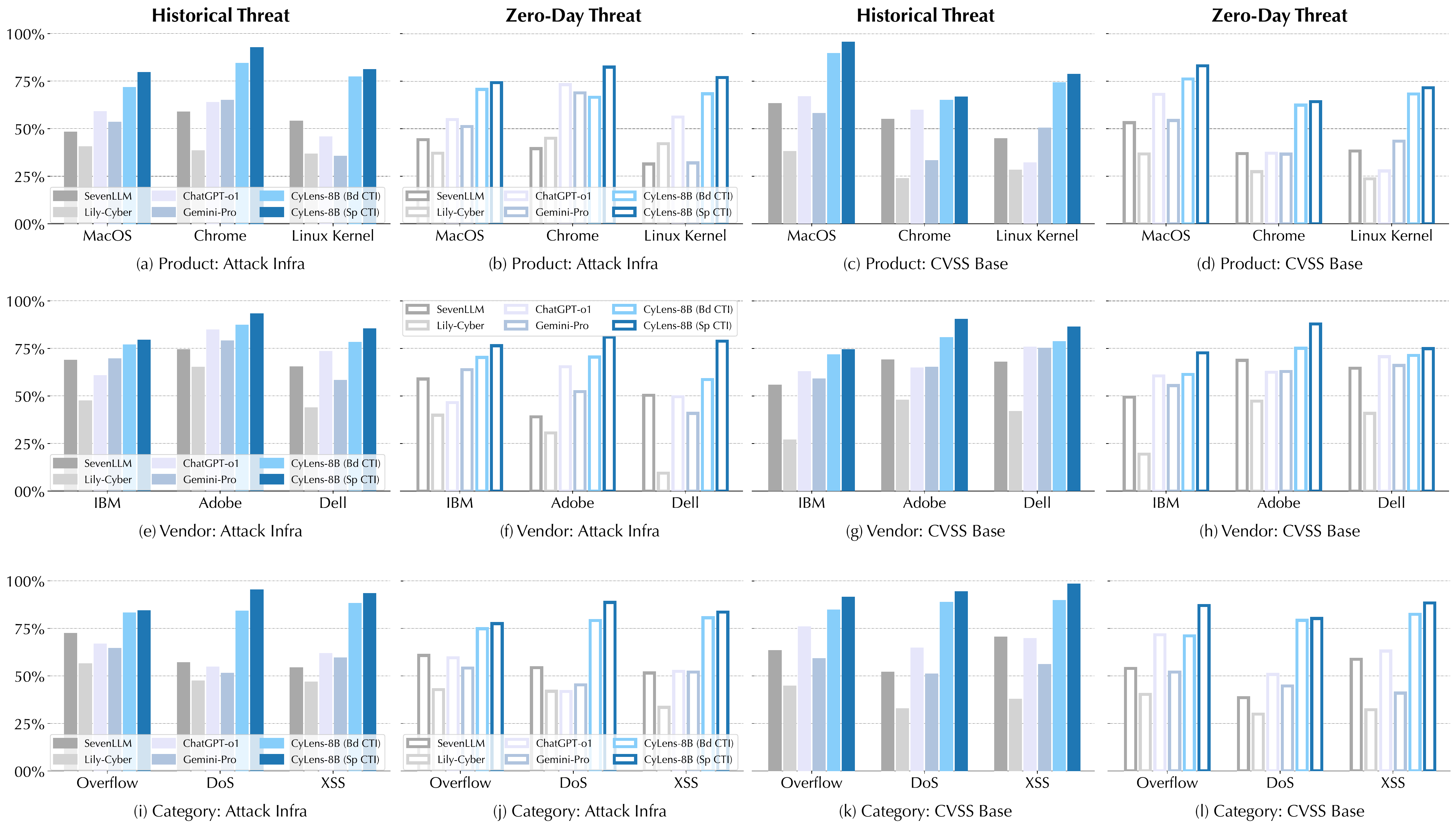}
    \caption{Additional evaluation results across on attack infrastructure contextualization (F1 score) and CVSS base severity classification (accuracy), complemented to Figure \ref{fig:organize}.}
    \label{fig:organize-2}
\end{figure*}

\subsection{Evaluation Data}
\label{app:eval-data}

{\bf Rationale.} Evaluating CTI using both historical cyber threats and zero-day threats provides a comprehensive assessment of their robustness and adaptability. Historical threats serve as a fundamental benchmark, allowing for the validation of defensive mechanisms against well-documented attack patterns, known TTPs. This retrospective analysis ensures that security measures remain effective against previously encountered threats and can detect variations of known exploits. On the other hand, zero-day threats introduce an element of unpredictability, testing the system's ability to detect and respond to novel attack vectors without predefined signatures or heuristics. 

By incorporating both types of threats, evaluations can measure not only the efficiency of existing threat detection models but also the adaptability of security solutions in dynamic and evolving threat landscapes. This dual approach is essential for developing resilient cybersecurity frameworks that balance historical insights with forward-looking threat intelligence.

\subsection{Evaluation Metrics}
\label{app:eval-metric}

We define the evaluation metrics used for each analytical target, along with the strategies and rationale behind their selection.

\begin{mitemize}
    \item {\bf Threat Actor:} Since threat actors are typically unique in each threat report, we use Accuracy to assess whether the threat actor generated by LLMs matches the reference. Additionally, we employ BERT Score to measure the textual similarity between the generated explanation and the reference for threat actor attribution. BERT Score is implemented using the Hugging Face library\footnote{https://huggingface.co/spaces/evaluate-metric/bertscore}.

    \item {\bf TTPs:} As TTPs are typically presented as a list, we adopt four metrics:
        \begin{mitemize}
            \item Precision – Measures the proportion of correctly identified TTPs among those predicted by the LLM.
            \item Recall – Measures the proportion of actual TTPs that were successfully identified.
            \item F1 Score – Provides a balanced measure of precision and recall.
            \item Intersection over Union (IoU) – Quantifies the overlap between the predicted and reference TTP sets.
        \end{mitemize}

    \item {\bf Campaign:} Since campaign is usually descriptive report, we use accuracy to have a binary judgment of whether LLM-generated campaign meets the reference. We also use BERT Score to measure text similarity between the generated and reference descriptions.

    \item {\bf Affected System \& Attack Infrastructure:} Similar to TTPs, both affected systems and attack infrastructures are structured as itemized lists. We therefore apply Precision, Recall, F1 Score, and IoU to assess the overlap between the LLM-generated results and the reference lists.

    \item {\bf Impact:} As impact analysis consists of descriptive statements, we use BERT Score to evaluate textual similarity between generated and reference statements.
 
    \item {\bf CVE and CWE Detection \& Correlation:}  Given the large-scale CVE candidates and hundreds of CWEs, the detection and correlation tasks prompt LLMs to generate the top-10 most relevant identifiers (CVE or CWE). We use Hit@10 to measure whether at least one correct identifier is present in the top-10 results.

    \item {\bf Correct Product \& Threat Category:} To assess whether the generated CVEs correspond to the correct affected product or fall within the correct threat category, we compute Accuracy as the evaluation metric.

    \item {\bf CVSS Metrics:} Since CVSS metrics have a limited number of categorical values (e.g., base severity levels), we use classification accuracy to evaluate the correctness of the predicted severity levels.

    \item {\bf EPSS Trend Prediction:} Since the EPSS score is a numerical probability (e.g., a 5.5\% chance of exploitation within a given timeframe), we calculate the Root Mean Square Error (RMSE) to measure the deviation between the LLM-predicted score and the ground truth.

    \item {\bf Tool Use \& Advisory Identification:} Both of these tasks involve selecting a single correct item, so we evaluate performance using Accuracy to measure whether the LLM can identify the correct tool or advisory.

    \item {\bf Code Patch \& Mitigation:} Since both tasks involve generating textual recommendations, we use BERT Score to assess the similarity between the LLM-generated output and the reference.
\end{mitemize}

\subsection{Additional Evaluation Results}
\label{app:eval-result}

Complementing \cref{sec:expt-adapt}, Figure \ref{fig:organize-2} presents additional evaluation results on specialized CTI tasks, focusing on attack infrastructure contextualization (measured by F1 score) and CVSS base severity classification (measured by accuracy). The results are reported across different products, vendors, and threat categories.

\end{document}